\documentstyle[onecolumn]{mn}
\newcommand{\mincir}{\raise
-2.truept\hbox{\rlap{\hbox{$\sim$}}\raise5.truept
\hbox{$<$}\ }}
\newcommand{\magcir}{\raise
-2.truept\hbox{\rlap{\hbox{$\sim$}}\raise5.truept
\hbox{$>$}\ }}
\newcommand{\minmag}{\raise-2.truept\hbox{\rlap{\hbox{$<$}}\raise
6.truept\hbox
{$>$}\ }}
\newcommand{\be}{\begin{equation}}
\newcommand{\ee}{\end{equation}}
\newcommand{\ba}{\begin{eqnarray}}
\newcommand{\ea}{\end{eqnarray}}
\newcommand{\brr}{\begin{array}}
 
\newcommand{\err}{\end{array}}
\newcommand{\bc}{\begin{center}}
\newcommand{\ec}{\end{center}}

\input{psfig.sty} 
\title{Predicting the Clustering of X-Ray Selected Galaxy Clusters in 
Flux-Limited Surveys}
\author[Moscardini et al.]
{L. Moscardini$^1$, S. Matarrese$^{2,3}$, F. Lucchin$^1$ and
P. Rosati$^4$ \\
$^1$Dipartimento di Astronomia, Universit\`a di Padova,
vicolo dell'Osservatorio 5, I--35122 Padova, Italy\\
$^2$Dipartimento di Fisica G. Galilei, 
Universit\`{a} di Padova, via Marzolo 8, I--35131 Padova, Italy\\
$^3$Max-Planck-Institut f\"ur Astrophysik, 
Karl-Schwarzschild-Strasse 1, D-85748 Garching, Germany \\
$^4$ESO -- European Southern Observatory, Karl-Schwarzschild-Strasse 2,
D--85748 Garching, Germany
}

\begin{document}

\maketitle

\begin{abstract}
We present a model to predict the clustering properties of X-ray
selected clusters in flux-limited surveys. Our technique correctly
accounts for past light-cone effects on the observed clustering and
follows the non-linear evolution in redshift of the underlying dark
matter correlation function and cluster bias factor.  The conversion
of the limiting flux of a survey into the corresponding minimum mass
of the hosting dark matter haloes is obtained by using theoretical and
empirical relations between mass, temperature and X-ray luminosity of
galaxy clusters. Finally, our model is calibrated to reproduce the
observed cluster counts adopting a temperature-luminosity relation
moderately evolving with redshift. We apply our technique to three
existing catalogues: the ROSAT Brightest Cluster sample (BCS); the
X-ray brightest Abell-type cluster sample (XBACs); the ROSAT-ESO Flux
Limited X-ray sample (REFLEX). Moreover, we consider an example of
possible future space missions with fainter limiting flux. In general,
we find that the amplitude of the spatial correlation function is a
decreasing function of the limiting flux and that the Einstein-de
Sitter models always give smaller correlation amplitudes than open or
flat models with low matter density parameter $\Omega_{\rm 0m}$. In
the case of the XBACs catalogue, the comparison with previous
estimates of the observational spatial correlation shows that only the
predictions of models with $\Omega_{\rm 0m}=0.3$ are in good agreement
with the data, while the Einstein-de Sitter models have too low a
correlation strength.  Finally, we use our technique to discuss the
best strategy for future surveys. Our results show that to study the
clustering properties of X-ray selected clusters the choice of a wide
area catalogue, even with a brighter limiting flux, is preferable to a
deeper, but with smaller area, survey.
\end{abstract}


\section{Introduction}

Extending the study of the matter distribution to the largest scales
reachable by observations can provide important constraints on models
for the formation of cosmic structures.  In fact, on very large scales
the present-day fluctuation field is just a linear amplification of
the primordial one. In the past years, surveys of galaxies have been
used to describe the spatial distribution of the cosmic structures up
to few hundred Mpc. It is now well established that an accurate and
efficient alternative way to describe the very large scale structure
of the universe is to use the spatial distribution of clusters of
galaxies. In the framework of the gravitational instability picture
galaxy clusters are the most extended gravitationally bound systems in
the universe. Moreover, their typical separation is much larger than
their expected displacements from the primordial positions. Therefore,
their study can be quite useful in putting constraints on the
cosmological parameters. This possibility is made easier by the fact
that the cluster clustering signal is enhanced with respect to the
galaxy one, because the clusters are expected to form in highly
overdense regions (peaks) of the cosmological density field and are
consequently strongly biased (Kaiser 1984).

For all these reasons, a large effort has been made to compile cluster
surveys, leading to extended redshift catalogues in the optical band
(see e.g. Postman 1998 for a review).  However, as first suggested by
Sutherland (1988), this kind of catalogues can be affected by strong
problems due to the spurious presence of interloper galaxies, which
would alter the general statistical properties of the clusters. This
problem does not affect surveys obtained in the X-ray band, where the
cluster emission, due to the thermal bremsstrahlung originated in the
hot intracluster plasma, is more concentrated around the centre,
because of its dependence on the square of the baryonic density.  In
the last twenty years, various space missions have been planned (and
launched) to build extended catalogues of X-ray selected clusters. To
this aim the role played by the ROSAT satellite has been quite
important. From its all-sky survey, which was carried out in the soft
(0.1--2.4 keV) X-ray band, different catalogues of clusters have been
built: the RASS1 Bright sample (De Grandi et al. 1999a), the BCS
sample (Ebeling et al. 1998), the XBACs sample (Ebeling et al. 1996)
and the REFLEX sample (B\"ohringer et al. 1998); more details about
these surveys can be found in Section 4.1.  These data have been
essentially used to compute the cluster number counts and the X-ray
luminosity function, which have relevant cosmological implications.
In particular, the analysis of the cluster abundance (also as a
function of redshift) has been largely used to provide estimates of
the mass fluctuation amplitude and of the matter density parameter
$\Omega_{\rm 0m}$ (e.g. Eke, Cole \& Frenk 1996; Viana \& Liddle 1996;
Mo, Jing \& White 1996; Oukbir, Bartlett \& Blanchard 1997; Eke et
al. 1998; Sadat, Blanchard \& Oubkir 1998; Viana \& Liddle 1999;
Borgani, Plionis \& Kolokotronis 1999; Borgani et al. 1999).

An alternative approach is based on the study of the spatial
distribution of X-ray selected clusters. The standard statistical
tools used to this aim are the (spatial and angular) two-point
correlation function and the power-spectrum.  Pioniering studies on
small samples have been performed by Lahav et al. (1989), Nichol,
Briel \& Henry (1994) and Romer et al. (1994), suggesting relatively
small values of the correlation length $r_0$. In particular Romer et
al. (1994) found $r_0 = 13-15\ h^{-1}$ Mpc ($h$ is the value of the
local Hubble constant $H_0$ in units of 100 km s$^{-1}$ Mpc$^{-1}$) by
analysing a sample of galaxy clusters selected from the ROSAT All-Sky
Survey.  Very recently the amount of data has become large enough to
allow more reliable estimates of the spatial correlation function
$\xi(r)$.  For example, Abadi, Lambas \& Muriel (1998) and Borgani,
Plionis \& Kolokotronis (1999) analysed the XBACs catalogue obtaining
correlation lengths in the range $r_0\approx 20-26\ h^{-1}$ Mpc, while
Moscardini et al. (2000) found $r_0= 21.5^{+3.4}_{-4.4}\ h^{-1}$ Mpc
using the RASS1 Bright Sample.  Notice that these values are larger
than the correlation amplitudes ($r_0\approx 13-18\ h^{-1}$ Mpc)
resulting from the optical data, once corrected for the previously
quoted projection effects (see e.g. the APM analysis performed by
Croft et al. 1997).  In the near future, with the new generation of
X-ray satellites, such as XMM and Chandra, the quality and quantity of
the cluster data will sensibly increase, giving the opportunity to
improve the clustering measurements and to better understand the
reasons for this difference.

In this paper we introduce a theoretical model to make predictions on
the correlations of X-ray selected clusters in flux-limited
surveys. The method, which fully accounts for the past light-cone
effects using a technique developped in Matarrese et al. (1997) and
Moscardini et al. (1998), also takes into account the non-linear
growth of clustering and the redshift evolution of the cluster bias
factor. The conversion of the limiting flux of a given survey into the
corresponding mass of the hosting dark matter haloes is made by using
theoretical and empirical relations between mass, temperature and
X-ray luminosity of galaxy clusters.  The same method described here
has been already applied in Moscardini et al. (2000), where a
comparison between observational results and theoretical predictions
for the RASS1 Bright Sample has been performed. A similar approach
(which, however, neglects the past light-cone effects and the
non-linear evolution of clustering) has been adopted by Borgani,
Plionis \& Kolokotronis (1999) in their analysis of the XBACs
catalogue. Suto et al. (2000) made quantitative predictions for future
surveys adopting a very similar method, using an equivalent formalism
to allow for past light-cone effects, but also including a model for
the redshift-space distortions.

The plan of the paper is as follows. In Section 2 we discuss our
method to study the clustering of a class of objects in the past
light-cone and present the relevant formulas for the (spatial and
angular) two-point correlation and for the power-spectrum. In Section
3 we introduce our theoretical model to estimate the correlations of
X-ray selected clusters. In particular, we discuss how to follow the
redshift evolution of the cluster bias and of the mass
auto-correlation function and how to convert the catalogue limiting
flux into a minimum halo mass. The different cosmological scenarios
here considered and the resulting X-ray cluster number counts are also
presented. Section 4 is devoted to the theoretical predictions of the
clustering properties for various present and future catalogues.
Section 5 presents a discussion of the robustness of the results with
respect to different choices of the parameters. Conclusions are drawn
in Section 6.

\section{Clustering in the past light-cone}

\subsection{Spatial two-point function}

Our aim is to obtain theoretical expectations for the correlation
properties of high-redshift objects like galaxies, clusters, etc. Let
us start by defining $n_{\rm obs}(r(z) \hat \gamma; z,M)$ as the
number of objects with redshift $z$ that an observer placed in the
origin measures in the angular direction specified by the unit vector
$\hat\gamma$, per unit comoving volume and per unit logarithmic
interval of some set of intrinsic properties (like mass, luminosity,
etc.), generally denoted by $M$.  Here $r(z)$ is the comoving radial
distance related to the redshift $z$ via the general law (the effect
of peculiar velocities on the redshift-distance relation is here
disregarded)
\be
r(z) = {c \over H_0 \sqrt{|\Omega_{0\cal R}|}} ~{\cal S}
\left(\sqrt{|\Omega_{0\cal R}|}
\int_0^z {dz'\over  E(z')} \right) \;,
\label{eq:x_z}
\ee
where (e.g. Peebles 1993)
\be
E(z)\equiv H(z)/H_0 = \left[\Omega_{\rm 0m}(1+z)^3 + \Omega_{0\cal R}(1+z)^2
+ \Omega_{0\Lambda}\right]^{1/2} \;,
\ee
and $\Omega_{0\cal R} \equiv 1 - \Omega_{\rm 0m} - \Omega_{0\Lambda}$
with $\Omega_{\rm 0m}$ and $\Omega_{0\Lambda}$ the present density
parameters for the non-relativistic matter and cosmological constant
components, respectively. In the above formula, for an open universe
model, $\Omega_{0\cal R}>0$, ${\cal S}(x)\equiv \sinh (x)$, for a
closed universe, $\Omega_{0\cal R}<0$, ${\cal S}(x)\equiv \sin (x)$,
while in the Einstein-de Sitter case, $\Omega_{0\cal R} = 0$, ${\cal
S}(x) \equiv x$.  In what follows we will also need the Jacobian
determinant
\be
g(z) \equiv  r^2(z) 
\biggl[1 + {H_0^2 \over c^2} ~\Omega_{0\cal R} ~r^2(z) \biggr]^{-1/2} 
{dr \over dz} \;.
\ee

To obtain an expression for the two-point correlation function, we
start by writing the average number of distinct pairs with relative
separation in the range $r, r+dr$, with redshift in the range ${\cal
Z}$ and $M$ in the domain ${\cal M}$. This reads
\ba
& \langle N_{pairs}(r) \rangle dr \ = &
{1\over 2} \int_{\cal M} d\ln M_1 d\ln M_2 
\int_{\cal Z} d z_1 d z_2 g(z_1) g(z_2)  \times \nonumber \\
& & \int_{4 \pi} d \Omega_{\gamma_1} 
d \Omega_{\gamma_2} \delta^D\bigl(r - |r_1 \hat \gamma_1 - r_2 \hat \gamma_2|
\bigr) \langle n_{\rm obs}(r(z_1) \hat \gamma_1; z_1,M_1)
n_{\rm obs}(r(z_2) \hat \gamma_2; z_2,M_2) \rangle dr \;,  
\label{eq:npairs0}
\ea
where $\delta^D$ is the Dirac delta function.  One can transform the
above delta function into a delta function over the relative angular
separation [accounting for the Jacobian $r/(r_1 r_2$)], integrate over
the angles and then proceed as in Matarrese et al. (1997) [see the
discussion after eq.(11)]. In such a calculation one can safely
neglect curvature corrections to the cosine rule, which would in
principle arise whenever $\Omega_{0\cal R} \neq 0$ (see in this
respect the discussion in Matarrese et al. 1997).  The result is the
exact expression
\be
\langle N_{pairs}(r) \rangle dr = {1\over 2} 
\int_{\cal Z} d z_1 d z_2 {{\cal N}(z_1) \over r(z_1)} {{\cal N}(z_2) \over 
r(z_2)} \biggl[ 1 + \xi_{\rm obj}(r;z_1,z_2) \biggr] r dr \;,
\label{eq:npairs}
\ee
where $\xi_{\rm obj}(r;z_1,z_2)$ is the theoretical correlation
function of the given objects at separation $r$ and redshifts $z_1,
z_2$, integrated over the domain of $M$ values. Here ${\cal N}(z)$ is
the actual redshift distribution of the catalogue, which is given by
${\cal N}(z) = \int_{\cal M} d \ln M {\cal N}(z,M)$, with ${\cal
N}(z,M) = 4\pi g(z) \phi(z,M) \bar n(z,M)$, where $\bar n(z,M)$ is the
expected number of objects per comoving volume at redshift $z$ and
$\phi(z,M)$ the isotropic catalogue selection function, which also
accounts for possible incomplete sky-coverage, as a function of
redshift and object intrinsic properties.

A somewhat simpler formula can be obtained by using the delta function
to integrate over one redshift and by making the following two {\it
approximations}: i) the redshift distribution ${\cal N}(z)$ is almost
constant over redshift intervals corresponding to the considered
comoving separation $r$, ii) the theoretical correlation does not
change considerably over the time intervals corresponding to the
separation $r$.  The resulting expression for the mean number of pairs
reads
\be
\langle N_{pairs}(r) \rangle dr = {1\over 2} 
\int_{\cal Z} d z {{\cal N}^2(z) \over g(z)} 
\biggl[ 1 + \xi_{\rm obj}(r;z) \biggr] r^2 dr \;. 
\label{eq:npairs2}
\ee

We therefore get two alternative expressions for the correlation
function: using eq.(\ref{eq:npairs}) above we find
\be
\xi_{\rm obs}(r) = { \int_{\cal Z} d z_1 d z_2 {\cal N}(z_1) r^{-1}(z_1)
{\cal N}(z_2) r^{-1}(z_2) ~\xi_{\rm obj}(r;z_1,z_2) \over 
\bigl[ \int_{\cal Z} d z_1 {\cal N}(z_1) r^{-1}(z_1) \bigr]^2 } \;,
\label{eq:xiobs}
\ee
[which differs from eq.(15) of Matarrese et al. (1997), by the extra
factors $1/r(z)$ in the redshift integrations], while using
eq.(\ref{eq:npairs2}) we obtain the approximate formula
\be
\xi_{\rm obs}(r) = {\int_{\cal Z} d z_1 {\cal N}^2(z_1) g^{-1}(z_1)
~\xi_{\rm obj}(r;z_1) \over \int_{\cal Z} d z_1 
{\cal N}^2(z_1) g^{-1}(z_1)} \;. 
\label{eq:xiobs2}
\ee
The latter expression has been independently derived by Yamamoto \&
Suto (1999).

Note that compared to eq.(15) of Matarrese et al. (1997) these
expressions give larger weight to lower redshift pairs. Nevertheless,
these pairs are also preferred by the selection function, so, one
might argue that the change implied is generally small.  Similar
conclusions have been reached by Yamamoto \& Suto (1999).  More in
general, one might introduce some weight $w(z)$ in the redshift
integrations in order to maximize the signal for the correlation
function, as suggested by Matsubara, Suto \& Szapudi (1997). In this
sense eq.(\ref{eq:xiobs}) above and eq.(15) in (Matarrese et
al. 1997), correspond to different choices for the weight.  In what
follows we will adopt the exact expression in eq.(\ref{eq:xiobs}) for
the correlation function. The problems related to the presence of a
double redshift dependence of the intrinsic object correlation in
eq.(\ref{eq:xiobs}) will be dealt with by replacing $z_1$ and $z_2$
with a suitable average redshift $z_{\rm ave}$. Following Porciani
(1997) we choose $z_{\rm ave}$ so that the linear growth factor of
density fluctuations $D_+(z_{\rm ave})$ equals the geometric mean of
$D_+(z_1)$ and $D_+(z_2)$; this is expected to give an accurate
approximation to the exact formula, even on mildly non-linear scales.

In our treatment we disregard the effect of redshift-space
distortions.  Analytical expressions have been obtained in the mildly
non-linear regime, by using either the Zel'dovich approximation
(Fisher \& Nusser 1996; Taylor \& Hamilton 1996; Hui, Kofman \&
Shandarin 1999) or higher order perturbation theory (Heavens,
Matarrese \& Verde 1998). The complicating role of the cosmological
redshift-space distortions on the evolution of the bias factor has
been considered by Suto et al. (1999,2000). Recently, Nishioka \&
Yamamoto (1999) have examined the redshift-space distortions effects
on the two-point correlation function and power-spectrum of
high-redshift objects; Magira, Jing \& Suto (2000) have extended the
formalism to account for the non-linear effects of the density and
velocity evolution. Suto et al. (2000), in a general study of the
two-point function of X-ray selected clusters, have included the
effect of both linear and non-linear redshift-space distortions. An
estimate of the effect of large-scale redshift-space distortions can
be obtained within linear theory and the distant-observer
approximation (Kaiser 1987; see Zaroubi \& Hoffman 1996 and Matsubara
1999 for an extension of this formalism to all-sky surveys). In this
case the enhancement of the redshift-space averaged power-spectrum is
given by the factor $1+2\beta(z)/3+\beta^2(z)/5$, where
$\beta(z)=f(z)/b_{\rm eff}(z)$.  In the previous expression $f\equiv
-d \ln D_+(z)/d \ln (1+z)\simeq \Omega_{\rm
m}^{0.6}(z)+\Omega_\Lambda(z)[1+\Omega_{\rm m}(z)/2]/70$ (Lahav et
al. 1991) and $b_{\rm eff}(z)$ is the effective bias (see
below). Plionis \& Kolokotronis (1998), by analysing the XBACs
catalogue and using linear perturbation theory to relate the X-ray
cluster dipole to the Local Group peculiar velocity, found $\beta
\simeq 0.24\pm 0.05$. Adopting this approach, Borgani, Plionis \&
Kolokotronis (1999) conclude that the overall effect of redshift-space
distortions is a small change of the correlation function, which
expressed in terms of $r_0$ corresponds to an $\simeq 8$ per cent
increase.  For deeper surveys, such as ABRIXAS (see below) the linear
redshift-space distortion becomes slightly smaller (approximately 6
per cent increase of the correlation length), because of the $b_{\rm
eff}$ increase with redshift.

\subsection{Power-spectrum}

The exact formula for the power-spectrum is given by the Fourier
transform of eq.(\ref{eq:xiobs}) above. However, an even simpler
expression can be obtained by Fourier transforming
eq.(\ref{eq:xiobs2}):
\be
P_{\rm obs}(k) = {\int_{\cal Z} d z {\cal N}^2(z) g^{-1}(z)
~P_{\rm obj}(k;z) \over \int_{\cal Z} d z {\cal N}^2(z) g^{-1}(z)} \;.
\label{eq:pow}
\ee
Quite recently, Yamamoto, Nishioka \& Suto (1999) have performed a
detailed and rigorous study of light-cone effects on the
power-spectrum.  It turns out that our expression in eq.(\ref{eq:pow})
reduces to the approximate formula in their eq.(16) on linear scales.
As Yamamoto, Nishioka \& Suto (1999) show, this simple expression is
accurate for wave-numbers $k \gg 1/r(z_{\rm max})$, where $r(z_{\rm
max})$ is the maximum redshift of the considered survey.

\subsection{Angular two-point function}

The angular correlation function $\omega_{\rm obs}(\vartheta)$ is
easily obtained in a similar way.  The number of pairs with relative
angular separation $\vartheta$ is defined similarly to
eq.(\ref{eq:npairs0}), where, however, the delta function over the
radial distances is replaced by one over directions in the sky, namely
$\delta^D ( \hat \gamma_1 \cdot \hat \gamma_2 - \cos\vartheta)$.
Integrating over all the angles one obtains
\be 
\omega_{\rm obs}(\vartheta) = N^{-2} 
\int_{\cal Z} d z_1 d z_2 
~{\cal N}(z_1) ~{\cal N}(z_2) 
\xi_{\rm obj}(r_{12};z_1,z_2) \;, \ \ \ \ 
r_{12} = \sqrt{r^2(z_1) + r^2(z_2) - 
2 r(z_1)r(z_2) \cos \vartheta} \;,
\ee
having once again neglected curvature corrections to the cosine rule.
Adopting then the {\em small-angle} approximation (e.g. Peebles 1980)
one gets the simpler expression
\be
\omega_{\rm obs}(\vartheta) = N^{-2} 
\int_{\cal Z} d z \biggl({dr \over dz}\biggr)^{-1}  
~{\cal N}^2(z) \int_{-\infty}^{\infty} du ~\xi_{\rm 
obj}[r(u,\vartheta,z),z] \;, 
\label{eq:ang}
\ee
with $r(u,\vartheta,z)=\sqrt{u^2 + r^2(z)\vartheta^2}$.  

\section{Modelling the two-point function of X-ray clusters}

\subsection{The effective bias of galaxy clusters}

To proceed in our modelling, we can safely assume that X-ray clusters
are in a one-to-one relation with virialized dark matter haloes and
take a linear bias model (see, however, Catelan et al. 1998 and
Catelan, Matarrese \& Porciani 1998, for a more refined bias
prescription), namely $\delta_{\rm cl}({\bf x}; M,z) \simeq
b(M,z)\delta_{\rm m}({\bf x},z)$, where $M$ is now the halo mass and
$z$ the considered redshift.  As a consequence we can write the object
two-point function as being proportional to the mass auto-correlation
function, namely $\xi_{\rm obj}(r;z_1,z_2) \approx b_{\rm eff}(z_1)
b_{\rm eff}(z_2) \xi(r,z_1,z_2)$. Here, following Matarrese et
al. (1997), we introduced the {\em effective bias} factor
\be 
b_{\rm eff}(z) \equiv {\cal N}^{-1}(z) \int_{\cal M} d\ln M' ~b(M',z) 
~{\cal N}(z,M')\, .
\label{eq:b_eff}
\ee
The assumption of a linear bias, as above, allows to further simplify
our expressions for correlation functions and power-spectrum. We get
\be 
\xi_{\rm obs}(r) = { \int_{\cal Z} d z_1 d z_2 
{\cal N}(z_1) r^{-1}(z_1) b_{\rm eff}(z_1)
{\cal N}(z_2) r^{-1}(z_2) b_{\rm eff}(z_2) ~\xi(r;z_{\rm ave}) \over 
\bigl[ \int_{\cal Z} d z_1 {\cal N}(z_1) r^{-1}(z_1) \bigr]^2 } \;,
\ee 
for the spatial two-point function,
\be
P_{\rm obs}(k) = {\int_{\cal Z} d z {\cal N}^2(z) g^{-1}(z) ~b^2_{\rm eff}(z)
~P(k;z) \over \int_{\cal Z} d z  {\cal N}^2(z) g^{-1}(z)} \;,
\label{eq:pow2}
\ee
for the power-spectrum, and
\be
\omega_{\rm obs}(\vartheta) = N^{-2} 
\int_{\cal Z} d z \biggl({dr \over dz}\biggr)^{-1}  
~{\cal N}^2(z) b^2_{\rm eff}(z) \int_{-\infty}^{\infty} 
du \xi[r(u,\vartheta,z),z] \;, 
\label{eq:ang2}
\ee
for the angular two-point function, which coincides with eq.(20) in
Matarrese et al. (1997).

In order to predict the clustering properties of our X-ray clusters as
a function of redshift we need to understand how the relation between
these objects and the underlying mass distribution evolves in time,
i.e. how the effective bias evolves.  For the cluster population it is
extremely reasonable to assume that structures on a given mass scale
are formed by the hierarchical merging of smaller mass units; for this
reason we can consider clusters as being fully characterized at each
redshift $z$ by the mass $M$ and formation epoch $z_f$ of their
hosting dark matter haloes.  For cluster-size haloes it is safe to
assume that instantaneous merging operates, so that $z_f=z$ [see,
e.g., the discussion in Kravtsov \& Klypin (1999); the effect of
taking $z\ne z_f$ in $\Omega_{\rm 0m}<1$ models has been discussed by
various authors (Kitayama \& Suto 1996; Viana \& Liddle 1996; Voit \&
Donahue 1998)] .  In this way their comoving mass function $\bar
n(z,M)$ can be computed using an approach derived from the
Press-Schechter (1974) technique. Moreover, it is possible to adopt
for the `monochromatic' bias $b(M,z)$ the expression which holds for
virialized dark matter haloes (e.g. Mo \& White 1996; Catelan et
al. 1998).  Recently, a number of authors (e.g. Sheth \& Tormen 1999
and references therein) have shown that the Press-Schechter relation
does not provide an accurate description of the halo abundance both in
the small-mass tail and in the large-mass one, which is more relevant
for the present study.  Also, the simple Mo \& White (1996) bias
formula has been shown not to correctly reproduce the correlation of
low mass haloes in numerical simulations. Several alternative fits
have been recently proposed (Jing 1998; Porciani, Catelan \& Lacey
1999; Sheth \& Tormen 1999; Jing 1999). In this paper we adopt the
relations recently introduced by Sheth \& Tormen (1999), which have
been shown to produce an accurate fit of the distribution of the halo
populations in the GIF simulations (Kauffmann et al. 1999).  The new
relations read
\be
\bar n(z,M) =  \sqrt{2 a A^2 \over \pi} \ 
{{3 H_0^2 \Omega_{\rm 0m}} \over {8\pi G}} \ 
{\delta_c 
\over M  D_+(z) \sigma_M } \ 
\biggl[ 1 + \biggl( {{D_+(z) \sigma_M} \over {\sqrt{a}  \delta_c}}
\biggr)^{2p} \biggr] \ 
  \bigg| {d \ln \sigma_M  \over d \ln M} \bigg| \  
\exp \biggl[ -{a \delta_c^2  \over 
2 D_+^2(z) \sigma^2_M}
\biggr] \; 
\label{eq:ps2}
\ee  
and
\be 
b(M,z) =   1 + {1 \over \delta_c }
\biggl( {a \delta_c^2 \over \sigma_M^2 D_+^2(z)} - 1\biggr) 
+ {2 p  \over \delta_c }
\biggl( {1 \over { 1+[\sqrt{a} \delta_c / (\sigma_M D_+(z))]^{2p}}}
\biggr) \; .
\label{eq:b_mono2}
\ee 
Here $\sigma^2_M$ is the mass-variance on scale $M$, linearly
extrapolated to the present time ($z=0$), $\delta_c$ the critical
linear overdensity for spherical collapse and $D_+(z)$ the linear
growth factor of density fluctuations, normalized to unity at $z=0$.
Following Sheth \& Tormen (1999), we adopt their best-fit parameters
$a=0.707$, $p=0.3$ and $A\approx 0.3222$, while the standard (Press \&
Schechter and Mo \& White) relations are recovered for $a=1$, $p=0$
and $A=1/2$.  Sheth, Mo \& Tormen (1999) have shown that these
expressions naturally arise when an ellipsoidal collapse model
replaces the usual spherical collapse in a Press-Schechter-like
approach.  A possible limitation with the use of this bias formula
comes from exclusion effects among dark matter haloes, which would
reduce the halo correlation function below some characteristic
separation, depending on the halo mass (e.g.  Sheth \& Lemson
1999). At first glance, one would expect this effect to take place
below the Lagrangian radius of the considered haloes. On the other
hand, as argued by Benson et al. (2000), haloes move somewhat from
their original position, so the largest exclusion effects should be
expected below the virial halo radius.  In our case, this limitation
would affect our estimate of the correlation function for separations
$r\ll 5 \ h^{-1}$ Mpc.

\subsection{Evolution of the mass auto-correlation function}

To predict the clustering properties of X-ray clusters we need a
description of the matter covariance function and its redshift
evolution.  To this purpose, Matarrese et al. (1997) and Moscardini et
al. (1998) used an accurate method, based on Hamilton et al. (1991)
original ansatz, as later developed by Peacock \& Dodds (1994), Jain,
Mo \& White (1995) and Peacock \& Dodds (1996), to evolve $\xi(r,z)$
into the fully non-linear regime. This technique allows to take into
account different background cosmologies and different initial
perturbation spectra, within the bottom-up hierarchical scenario for
structure formation in the Universe.

We adopt here the method of Peacock \& Dodds (1996), which deals with
the (dimensionless) power-spectrum $\Delta^2$:
\be
\Delta^2(k,z) = {1\over {2\pi^2}} k^3 P(k,z)\ ,
\ee
which is related to the two-point correlation function by
\be
\xi(r,z) = \int {{dk}\over{k}} \Delta^2(k,z) {{\sin kr}\over{kr}} \;. 
\ee
In the linear regime, one has $\Delta^2_{\rm lin}(k,z) = D_+^2(z)
\Delta^2_{\rm lin}(k,0)$. According to Peacock \& Dodds (1996) the 
non-linear power-spectrum is related to the linear one through the
transformation
\be
\Delta^2(k,z) = {\cal F}[\Delta^2_{\rm lin}(k_0,z)]\;, \ \
\ \ \ \
k_0=[1+\Delta^2(k,z)]^{-1/3} k \;,
\ee
where $k_0$ and $k$ are the linear and non-linear wavenumbers,
respectively, and the specific form of the fitting function ${\cal
F}$, whose detailed form is given in Peacock \& Dodds (1996), depends
on the growth suppression factor $g_\delta(z) \equiv (1+z) D_+(z)$. In
the general case $g_\delta$ contains a dependence on the background
cosmology. Carroll, Press \& Turner (1992) found the approximate (but
almost exact) expression
\be
g_\delta(z|\Omega_{\rm m},\Omega_\Lambda) = {5\over 2} 
\Omega_{\rm m} [\Omega_{\rm m}^{4/7}-\Omega_\Lambda + 
(1+\Omega_{\rm m}/2)(1+\Omega_\Lambda/70)]^{-1}\;,
\ee
with $\Omega_{\rm m} = \Omega_{\rm 0m} (1+z)^3/E^2(z)$ and
$\Omega_\Lambda = \Omega_{0\Lambda}/E^2(z)$.

The form of ${\cal F}(x)$ given by Peacock \& Dodds (1996) assumes a
power-law initial spectrum described by an index $n$. For models which
are not described by pure power-law spectra, such as the Cold Dark
Matter (CDM) models adopted here, one can use the same formulas, but
replacing $n$ by an effective index $n_{\rm eff}$, defined by
\be
n_{\rm eff}(k_0,z) = \frac{d \ln P_{\rm lin}(k,z)}
{d \ln k} \bigg|_{k=k_0(z)/2}\ .
\ee
According to Peacock \& Dodds (1996) this prescription is able to
reproduce the non-linear evolution with a precision of few per cent
(see also the discussion in Kravtsov \& Klypin 1999).  The conditions
under which this formalism can be applied to CDM models with $n<1$ are
discussed by Moscardini et al. (1998).

\subsection{Structure formation models}

In the following analysis we consider five models, all normalized to
reproduce the local cluster abundance, following the Eke, Cole \&
Frenk (1996) analysis of the temperature distribution of X-ray
clusters (Henry \& Arnaud 1991). All of them belong to the general
class of CDM models; their linear power-spectrum can be represented as
$P_{\rm lin}(k,0) \propto k^n T^2(k)$, where, for the CDM transfer
function $T(k)$, we use the Bardeen et al. (1986) fit. In particular,
we consider three different Einstein-de Sitter models, for which the
power-spectrum amplitude corresponds to $\sigma_8=0.52$ (here
$\sigma_8$ is the r.m.s. fluctuation amplitude in a sphere of $8
h^{-1}$ Mpc). They are: a version of the standard CDM (SCDM) model
with shape parameter (see its definition in Sugiyama 1995)
$\Gamma=0.45$ and spectral index $n=1$; the so-called $\tau$CDM model
(White, Gelmini \& Silk 1995), with $\Gamma=0.21$ and $n=1$; a tilted
model (hereafter TCDM; Lucchin \& Matarrese 1985), with $n=0.8$ and
$\Gamma=0.41$, corresponding to a high (10 per cent) baryonic content
(e.g. White et al.  1996; Gheller, Pantano \& Moscardini 1998). We
also consider an open CDM model (OCDM), with matter density parameter
$\Omega_{\rm 0m}=0.3$ and $\sigma_8=0.87$ and a low-density flat CDM
model ($\Lambda$CDM), with $\Omega_{\rm 0m}=0.3$, with $\sigma_8=0.93$
(see e.g. Liddle et al. 1996a,b and references therein).  Except for
SCDM, which is shown as a reference model, all these models are also
consistent with the level of fluctuations observed by COBE (Bunn \&
White 1997); for TCDM consistency is achieved by taking into account
the possible contribution of gravitational waves to large-angle CMB
anisotropies (e.g. Lucchin, Matarrese \& Mollerach 1992; Lidsey \&
Coles 1992).  A summary of the parameters of the cosmological models
used in this paper is given in Table \ref{t:models}.

\begin{table}
\centering
\caption[]{The parameters of the cosmological models. Column 2: the present
matter density parameter $\Omega_{\rm 0m}$; Column 3: the present
cosmological constant contribution to the density $\Omega_{0\Lambda}$;
Column 4: the primordial spectral index $n$; Column 5: the Hubble
parameter $h$; Column 6: the shape parameter $\Gamma$; Column 7: the
spectrum normalization $\sigma_8$; Column 8: the value of the
parameter $\eta$ in the temperature-luminosity relation required to
reproduced the observed $\log N$--$\log S$ (see text for details).}
\tabcolsep 4pt
\begin{tabular}{lccccccc} \\ \\ \hline \hline
Model & $\Omega_{\rm 0m}$ & $\Omega_{0\Lambda}$ & $n$ & $h$ &
$\Gamma$ & $\sigma_8$ & $\eta$  \\ \hline
SCDM         & 1.0 & 0.0 & 1.0 & 0.50 & 0.45 & 0.52 & -0.8 \\
$\tau$CDM    & 1.0 & 0.0 & 1.0 & 0.50 & 0.21 & 0.52 &  0.0 \\
TCDM         & 1.0 & 0.0 & 0.8 & 0.50 & 0.41 & 0.52 & -0.3 \\
OCDM         & 0.3 & 0.0 & 1.0 & 0.65 & 0.21 & 0.87 & -0.3 \\
$\Lambda$CDM & 0.3 & 0.7 & 1.0 & 0.65 & 0.21 & 0.93 & -0.2 \\
\hline
\end{tabular}
\label{t:models}
\end{table}

\subsection{From the catalogue limiting flux to the halo mass}

In order to predict the abundance and clustering of X-ray clusters in
a given sample we need to relate the X-ray cluster fluxes to a
corresponding halo mass at each redshift. The given band flux $S$
corresponds to an X-ray luminosity
\be
L_X(z,S) = 4\pi d_L^2(z) S \ ,
\label{eq:lx-s}
\ee
where $d_L=(1+z) r(z)$ is the luminosity distance. To convert $L_X$
into the total luminosity $L_{\rm bol}$ we perform band and bolometric
corrections by means of a Raymond-Smith code, where an overall ICM
metallicity of $0.3$ times solar is assumed (see e.g. Borgani et
al. 1999). We translate the cluster bolometric luminosity into a
temperature, adopting the empirical relation
\be
T = {\cal A} \ L_{\rm bol}^{\cal B} \ (1+z)^{-\eta} \;,
\label{eq:t-l}
\ee
where the temperature is expressed in keV and $L_{\rm bol}$ is in
units of $10^{44} h^{-2}$ erg s$^{-1}$. In the following analysis we
assume ${\cal A}=4.2$ and ${\cal B}=1/3$; these values allow a good
representation of the local data for temperatures larger than $\approx
1$ keV (e.g. David et al. 1993; White, Jones \& Forman 1997;
Markevitch 1998).  Analysing a catalogue of local compact groups,
Ponman et al. (1996) showed that at lower temperatures the $T-L_{\rm
bol}$ relation has a steeper slope (${\cal B}\approx 0.1$). For these
reasons we prefer to fix a minimum value for the temperature at $T=1$
keV. Moreover, even if observational data are consistent with no
evolution in the $T-L_{\rm bol}$ relation out to $z \approx 0.4$
(Mushotzky \& Scharf 1997; Donahue et al. 1999), a moderate redshift
evolution described by the parameter $\eta$ has been introduced to
reproduce the observed $\log N$--$\log S$ relation in the range $2
\times 10^{-14} \le S \le 2 \times 10^{-11}$ (see below).  A similar
approach has been followed also by Kitayama \& Suto (1997), Mathiesen
\& Evrard (1998) and Borgani et al. (1999).

Finally, with the standard assumption of virial isothermal gas
distribution and spherical collapse, it is possible to convert the
cluster temperature into the mass of the hosting dark matter halo,
namely (e.g. Eke, Cole \& Frenk 1996)
\be
T = {7.75 \over \beta_{\rm TM}} {\left(M\over {10^{15} h^{-1} 
M_\odot}\right)}^{2/3} 
E^{2/3}(z)
\left({\Delta_{\rm vir}(z) \over {178}}\right)^{1/3} \ .
\label{eq:t-m}
\ee
The quantity $\Delta_{\rm vir}$ represents the mean density of the
virialized halo in units of the critical density at that redshift
(e.g. Bryan \& Norman 1998 for fitting formulas). We assume
$\beta_{\rm TM}=1.17$, which is in agreement with the results of
different hydrodynamical simulations (Bryan \& Norman 1998; Gheller,
Pantano \& Moscardini 1998; Frenk et al. 1999).  Voit \& Donahue
(1998) discussed the validity of the previous relation in the case of
small $\Omega_{\rm 0m}$, where the assumption of a correspondence
between the cluster formation redshift and that at which we are
observing it is less accurate.

Once the relation between observed flux and halo mass at each redshift
is established we can obtain the redshift distribution ${\cal N}(z)$
as
\be
{\cal N}(z) = 4\pi g(z) \int_{\cal M} d \ln M  \phi(z,M) \bar n(z,M) \ ,
\label{eq:nz}
\ee
where the selection function $\phi(z,M)$ accounts for the sample sky
coverage $\Omega_{\rm sky}(S)$, which is formally defined as the area
of the sky covered by the sample as a function of the limiting flux
$S$, i.e.  $\phi(z,M) = \Omega_{\rm sky}[S(z,M)]/ 4 \pi$.

\begin{figure*}
\centering  
\psfig{figure=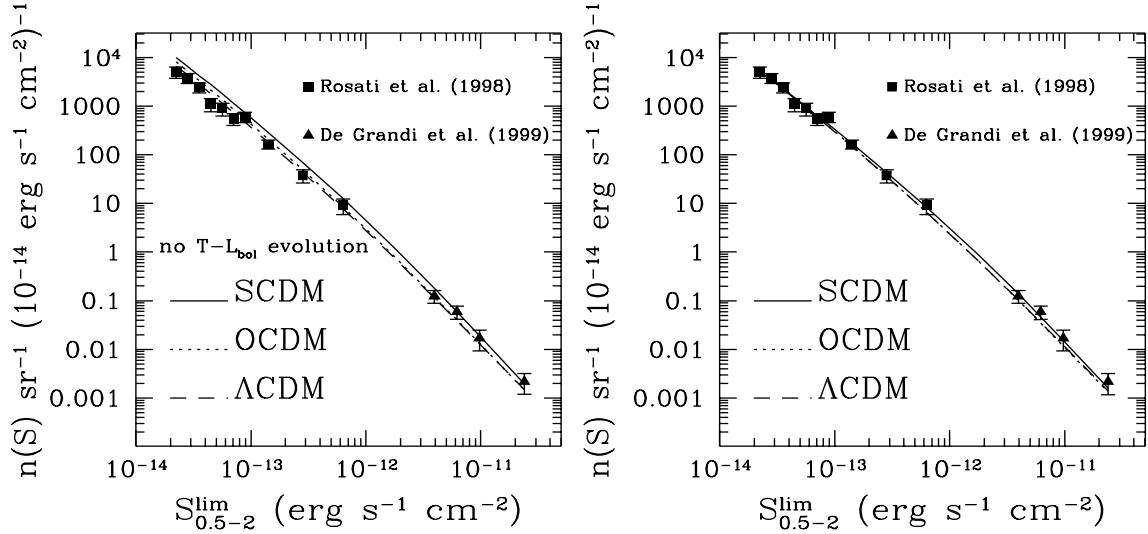,height=8.cm,width=16cm,angle=0}
\caption{The differential number counts (per unit solid angle) 
$n(S)$ as a function of the limiting flux $S_{\rm lim}$ computed in
the 0.5--2 keV band.  The observational results (with 1-$\sigma$
errorbars) obtained by Rosati et al. (1998) and De Grandi et
al. (1999a) are shown by filled squares and filled triangles,
respectively.  Different lines refer to theoretical predictions for
various cosmological models: SCDM (solid lines), OCDM (dotted lines)
and $\Lambda$CDM (dashed lines).  The left panel shows the results
obtained with no evolution in the temperature-luminosity relation
[i.e. $\eta=0$ in eq.(\ref{eq:t-l})]; the right panel presents the
results for the models adopting the value of $\eta$ which best-fits
the observational data (see text for more details).}
\label{fi:ns}
\end{figure*}

In Figure~\ref{fi:ns} we show the differential number counts $n(S)$
(per unit solid angle) as a function of the limiting flux $S_{\rm
lim}$ (defined in the 0.5--2 keV band), for three different
cosmological models (SCDM, OCDM and $\Lambda$CDM). The differential
number counts are computed from the relation
\be
n(S) = \int_0^\infty dz  ~g(z) ~\bar n(z,M) ~{{\partial \ln M}\over 
{\partial S}}\bigg|_z \;, 
\ee
assuming $\phi=1$.  In the same plot we report the observational data
coming from the RDCS sample (Rosati et al. 1998) up to fluxes of
$\simeq 5\times 10^{-13}$ erg s$^{-1}$ cm$^{-2}$ and from the RASS1
catalogue (De Grandi et al. 1999a) for higher fluxes.  The left panel
shows the theoretical predictions obtained under the assumption of no
evolution in the temperature-luminosity relation, i.e. with $\eta=0$
in eq.(\ref{eq:t-l}).  The agreement with the observational data is
good but there is some tendency to overestimate the number counts at
very low fluxes. The situation is improved in the right panel which
shows the results obtained allowing a redshift evolution of the
$T-L_{\rm bol}$ relation to fit the data. In this case the observed
$\log N-\log S$ relation is well reproduced by all cosmological
models. The required best-fitting values of $\eta$ for SCDM,
$\tau$CDM, TCDM, OCDM and $\Lambda$CDM are reported in Table
\ref{t:models}.  In the following analysis we will show results
obtained with these values of $\eta$; a short discussion of the effect
of the alternative choice $\eta=0$ will be presented in Section 5.1.

We note that the SCDM model predicts more clusters than low-density
models. This might appear counter-intuitive. In fact the cosmological
models are normalized using the local cluster abundance which
declines, when the matter density is high, more rapidly with
increasing redshift. This effect (shown also in the following redshift
distributions) is due to the larger number of low-temperature clusters
(we assume a minimum temperature of 1 keV) predicted by the SCDM
model. We remind that the cluster abundance normalisation results from
the analysis of objects with a typical temperature of about 5-6
keV. The use of a different normalization can help in sorting this
problem out.  The point is discussed in Section 5.2, where we show how
our results change if $\sigma_8$ is desumed from the X-ray luminosity
function which extends to clusters with temperature down to $\sim 1$
keV.
 
\section{Theoretical predictions for various catalogues}

\subsection{Description of the  catalogues}

In the following analysis we will apply our method to four different
cases. The first three applications refer to presently existing data:
the ROSAT Brightest Cluster Sample (BCS); the X-ray brightest
Abell-type cluster sample (XBACs); the ROSAT-ESO Flux Limited X-ray
sample (REFLEX). The fourth and last case will be an example of
possible future space missions: for that we will consider a survey
with characteristics similar to those which were expected from the
unfortunate satellite ABRIXAS. Here we will give the relevant
information about these surveys. We refer to the original papers for
more details.

\begin{itemize}
\item
The BCS catalogue (Ebeling et al. 1997, 1998; Crawford et al. 1999) is
an X-ray selected, flux-limited sample of 201 galaxy clusters with
$z\le 0.3$ drawn from the ROSAT All-Sky Survey in the northern
hemisphere ($\delta \ge 0^o$) and at high Galactic latitudes
($|b_{II}|\ge 20^o$).  The limiting flux is $S_{\rm lim}=4.45 \times
10^{-12}$ erg cm$^{-2}$ s$^{-1}$ in the 0.1--2.4 keV band. Since its
sky-coverage $\Omega_{\rm sky}(S)$ is not available, we use
$\Omega_{\rm sky}(S) = {\rm const} \simeq 4.13 $ steradians for fluxes
larger than $S_{\rm lim}$.

\item
The XBACs catalogue (Ebeling et al. 1996) is an all-sky X-ray sample
of 242 Abell galaxy clusters extracted from the ROSAT All-Sky Survey
data. Being optically selected, it is not a complete flux-limited
catalogue.  The sample covers high Galactic latitudes ($|b_{II}|\ge
20^o$). The adopted limiting flux is $S_{\rm lim}=5 \times 10^{-12}$
erg cm$^{-2}$ s$^{-1}$ in the 0.1--2.4 keV band.  Also in this case,
since the actual sky coverage is not published, we will adopt
$\Omega_{\rm sky}(S)= {\rm const} \simeq 8.27 $ steradians for fluxes
larger than $S_{\rm lim}$. Due to the aforementioned selection
effects, the XBACs luminosity function $N(L)$ in the faint part is
much lower than that obtained from other catalogues (e.g.  Ebeling et
al. 1997; Rosati et al. 1998; De Grandi et al. 1999b). Using a
redshift evolution of the temperature-luminosity relation, we forced
our models to be consistent with the number counts. For this reason we
have to introduce in the models for XBACs its incompleteness $I(L)$,
defined as the ratio between its luminosity function $N_{\rm
XBACs}(L)$ and $N_{\rm best}(L)$, which is a combination of the
results for RDCS at low $L$ (Rosati et al. 1998) and for BCS at high
$L$ (Ebeling et al. 1997):
\be
I(L)=N_{\rm XBACs}(L)/N_{\rm best}(L)\ .
\ee
The adopted parameters for the luminosity function, usually fitted as
$N(L)=K L^{-\alpha} \exp(L/L^*)$, are $K=2.8$, $\alpha=1.1$, $L^*=5.5$
for $N_{\rm XBACs}$ and $K=3.26$, $\alpha=1.83$, $L^*=5.5$ for $N_{\rm
best}$; in the previous formula $L^*$ is in units of $10^{44}$ erg
s$^{-1}$ in the band 0.5--2 keV (with $h=0.5$) and $K$ is in units of
$10^{-7}$ Mpc$^{-3}$ $L^{\alpha-1}$. The clustering properties of this
catalogue have been studied by different authors (Abadi, Lambas \&
Muriel 1998; Borgani, Plionis \& Kolokotronis 1999) giving a
correlation length in the range $ 20 \mincir r_0 \mincir 26 h^{-1}$
Mpc.

\item
The REFLEX survey (B\"ohringer et al. 1998; Guzzo et al. 1999) is a
large sample of optically confirmed X-ray clusters selected from the
ROSAT All-Sky Survey.  The sample, nearly completed, will contain
about 700 clusters in the southern hemisphere, at high Galactic
latitude ($|b_{II}|\ge 20^o$).  For our computations, we use the
actual sky coverage kindly provided by H. B\"ohringer and C. Collins
and defined in the ROSAT band (0.1--2.4 keV). In order to make
predictions for the cluster sample analysed by Collins et al. (2000),
which contains approximately 450 objects, we adopt the sky coverage
only above a minimum flux of $S_{\rm lim}=3 \times 10^{-12}$ erg
cm$^{-2}$ s$^{-1}$, where it falls to 97.3 per cent of the whole
surveyed region (4.24 steradians).  The clustering properties of a
part of this catalogue, known as RASS1 Bright Sample (De Grandi et
al. 1999a) and containing 130 galaxy clusters with flux larger than
$S_{\rm lim}=3 \times 10^{-12}$ erg cm$^{-2}$ s$^{-1}$ in the ROSAT
soft band 0.5--2 keV, have been analysed by Moscardini et al. (2000)
and compared with the predictions of different cosmological models
obtained with the same technique presented here.

\item
The ABRIXAS satellite (Tr\"umper, Hasinger \& Staubert 1998) has been
unluckily lost at the end of April 1999 because of problems with
energy supply. We use here the characteristics of the survey of X-ray
selected clusters which was expected to be obtained from its
observations as an example of an application of our method to possible
future samples of X-ray galaxy clusters. In the plans, the ABRIXAS
catalogue would have covered the area at high Galactic latitudes
($|b_{II}|\ge 20^o$) up to a limiting flux of $S_{\rm lim}=5 \times
10^{-13}$ erg cm$^{-2}$ s$^{-1}$ in the 0.5--2 keV band. We will
assume a constant sky coverage $\Omega_{\rm sky}(S) \simeq 8.27 $
steradians for fluxes larger than $S_{\rm lim}$.
\end{itemize}

\subsection{Physical properties of the clusters in different catalogues}

In this subsection we discuss the redshift dependence of the physical
properties of the clusters contained in the catalogues presented in
the previous subsection. These results are obtained using the
relations described in Section 3.4 and linking the limiting flux to
the X-ray luminosity, the luminosity to the temperature and finally
the temperature to the mass of the hosting dark matter halo.

\begin{figure*}
\centering  
\psfig{figure=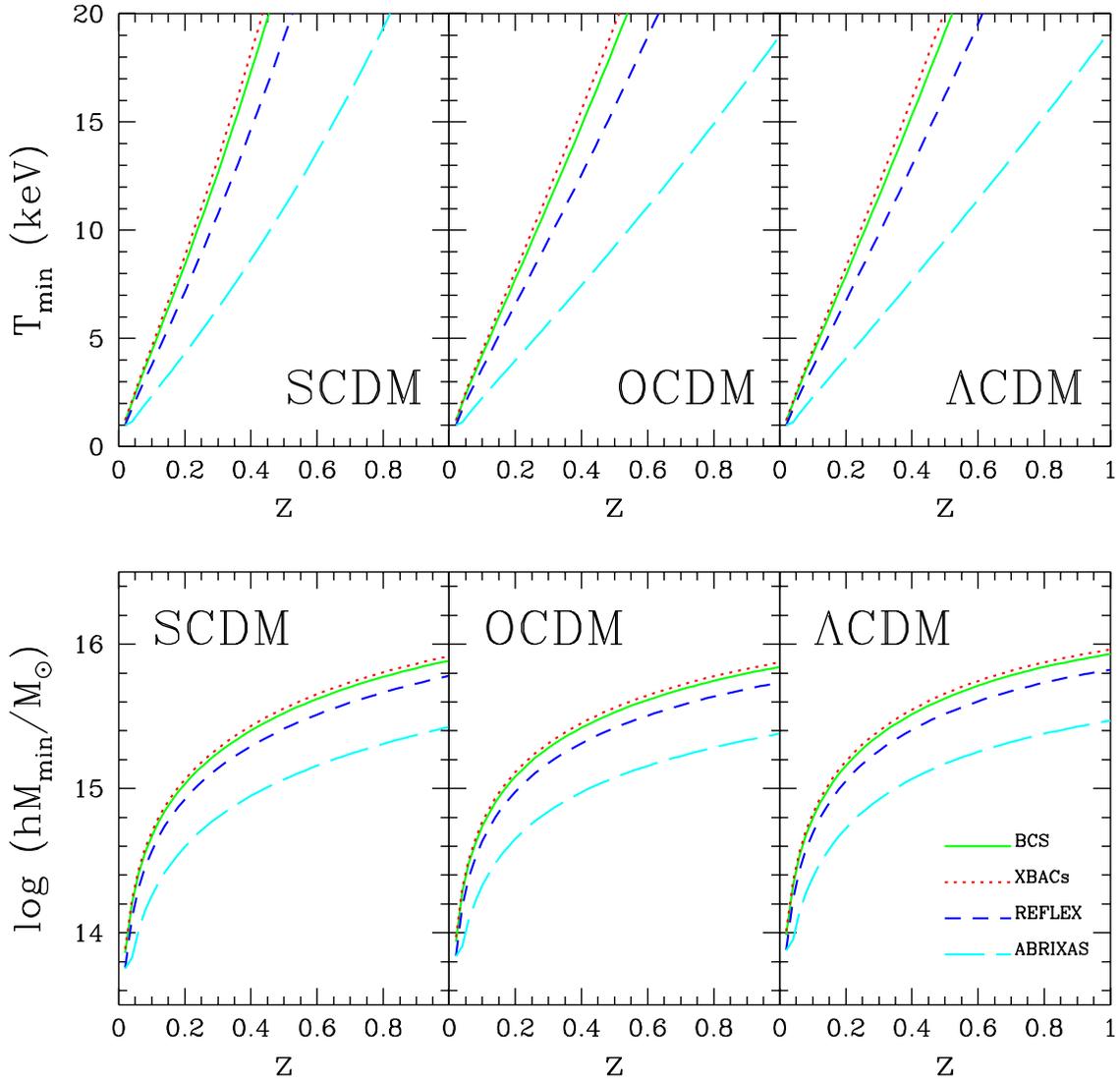,height=16.cm,width=16cm,angle=0}
\caption{The behaviour of the minimum temperature $T_{\rm min}$ (upper
panels) and of the minimum mass $M_{\rm min}$ (lower panels) for the
clusters included in the flux-limited samples as a function of the
redshift $z$. Different lines refer to results for different
catalogues: BCS (solid line), XBACs (dotted line), REFLEX (short
dashed) and ABRIXAS (long dashed). Predictions for different
cosmological models are shown in the different columns: SCDM (left),
OCDM (centre) and $\Lambda$CDM (right).}
\label{fi:tempmass}
\end{figure*}

In the upper panels of Figure~\ref{fi:tempmass} we present the
behaviour (as a function of the redshift $z$) of the minimum
temperature $T_{\rm min}$ corresponding to the limiting flux of the
various catalogues. The results are here presented only for three
cosmological models (SCDM, OCDM, $\Lambda$CDM). The minimum
temperature is a strongly increasing function of both the redshift and
the limiting flux, while the dependence on the cosmological model is
not so evident. For example, clusters with a temperature as high as 10
keV can enter the catalogues only up to $z\simeq 0.2$ when the
limiting flux of BCS and XBACs is considered, $z\simeq 0.3-0.35$ for
the REFLEX limits, and $z\simeq 0.4-0.5$ for the ABRIXAS ones.
 
In the lower panels of Figure~\ref{fi:tempmass} we show the redshift
dependence of the minimum mass $M_{\rm min}$. Once again the result is
strongly dependent on $S_{\rm lim}$. As a consequence, the different
samples can contain clusters with quite different ranges of masses:
given a redshift, the BCS and XBACs catalogues (which have very
similar limits, therefore leading to very similar minimum temperatures
and masses) tend to have richer (more massive) clusters than the
ABRIXAS and REFLEX ones.  Note that Suto et al. (2000) found that
$M_{\rm min}$ corresponding to a given value of the limiting flux
$S_{\rm lim}$ decreases for $z \magcir 1$. This effect is expected due
to the redshift dependences in eqs.(\ref{eq:lx-s}-\ref{eq:t-m}). Our
different choices for ${\cal B}$ and $\eta$ in eq.(\ref{eq:t-l}) shift
the turn-around of $M_{\rm min}$ to much larger redshifts, not
relevant for this study.

\begin{figure*}
\centering  
\psfig{figure=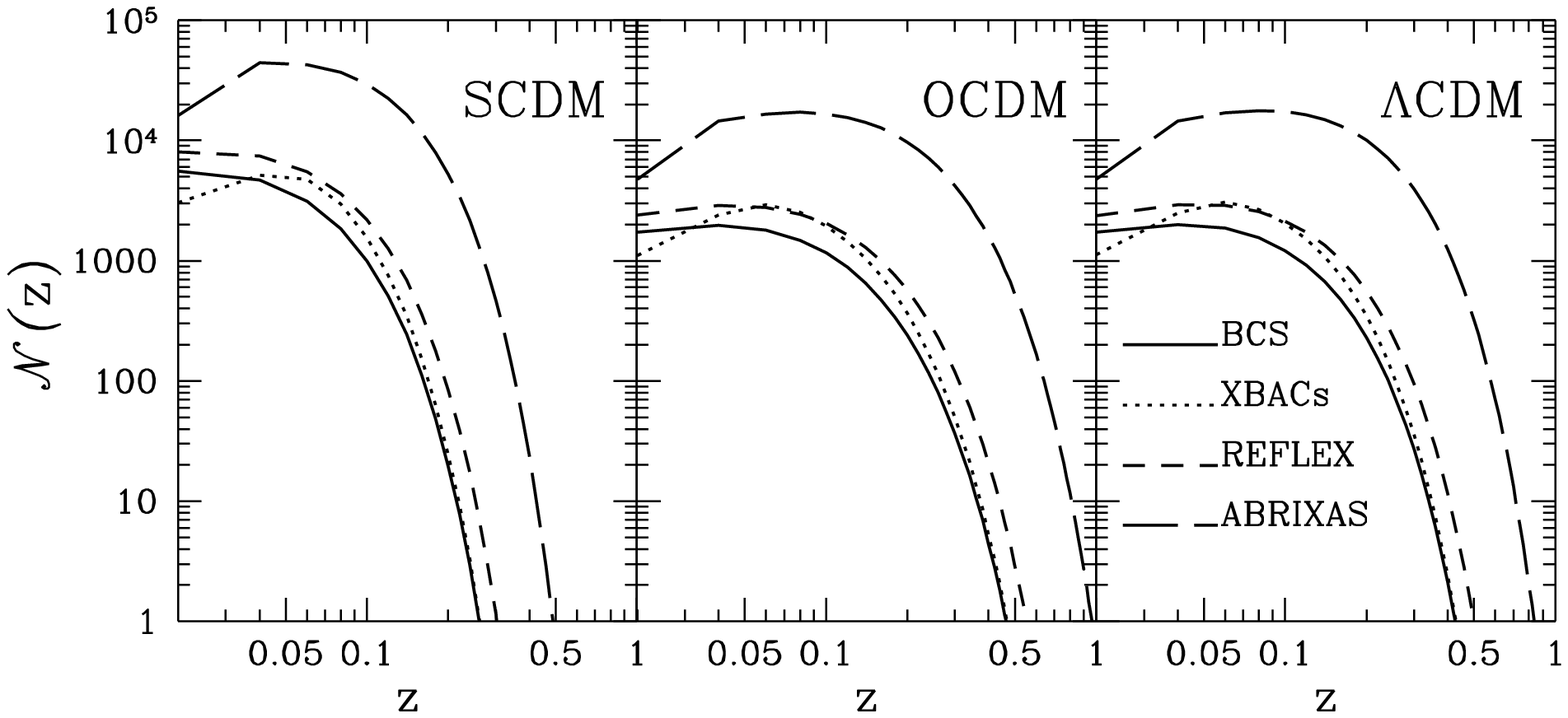,height=8.cm,width=16cm,angle=0}
\caption{The redshift distribution ${\cal N}(z)$ for the clusters included  
in the flux-limited samples.  Different lines refer to results for
different catalogues: BCS (solid line), XBACs (dotted line), REFLEX
(short dashed) and ABRIXAS (long dashed). Different panels show the
predictions for various cosmological models: SCDM (left), OCDM
(centre) and $\Lambda$CDM (right).  }
\label{fi:nz}
\end{figure*}

In order to predict the clustering properties, in our model one needs
to know the expected redshift distribution ${\cal N}(z)$ for the given
catalogue.  The results, computed by using eq.(\ref{eq:nz}), are shown
in Figure~\ref{fi:nz} for SCDM, OCDM and $\Lambda$CDM models. Of
course, the number of clusters increases with decreasing limiting
flux. In this case the differences between BCS and XBACs, which have
similar $S_{\rm lim}$, are due to the different sky coverage and to
the function $I(L)$ introduced to correct for the XBACs'
incompleteness.  As expected, the redshift distribution for the
Einstein-de Sitter model is less extended towards high redshifts than
in the models with low matter density parameter, due to the freezing
of the perturbation growth in the latter case.

\begin{figure*}
\centering  
\psfig{figure=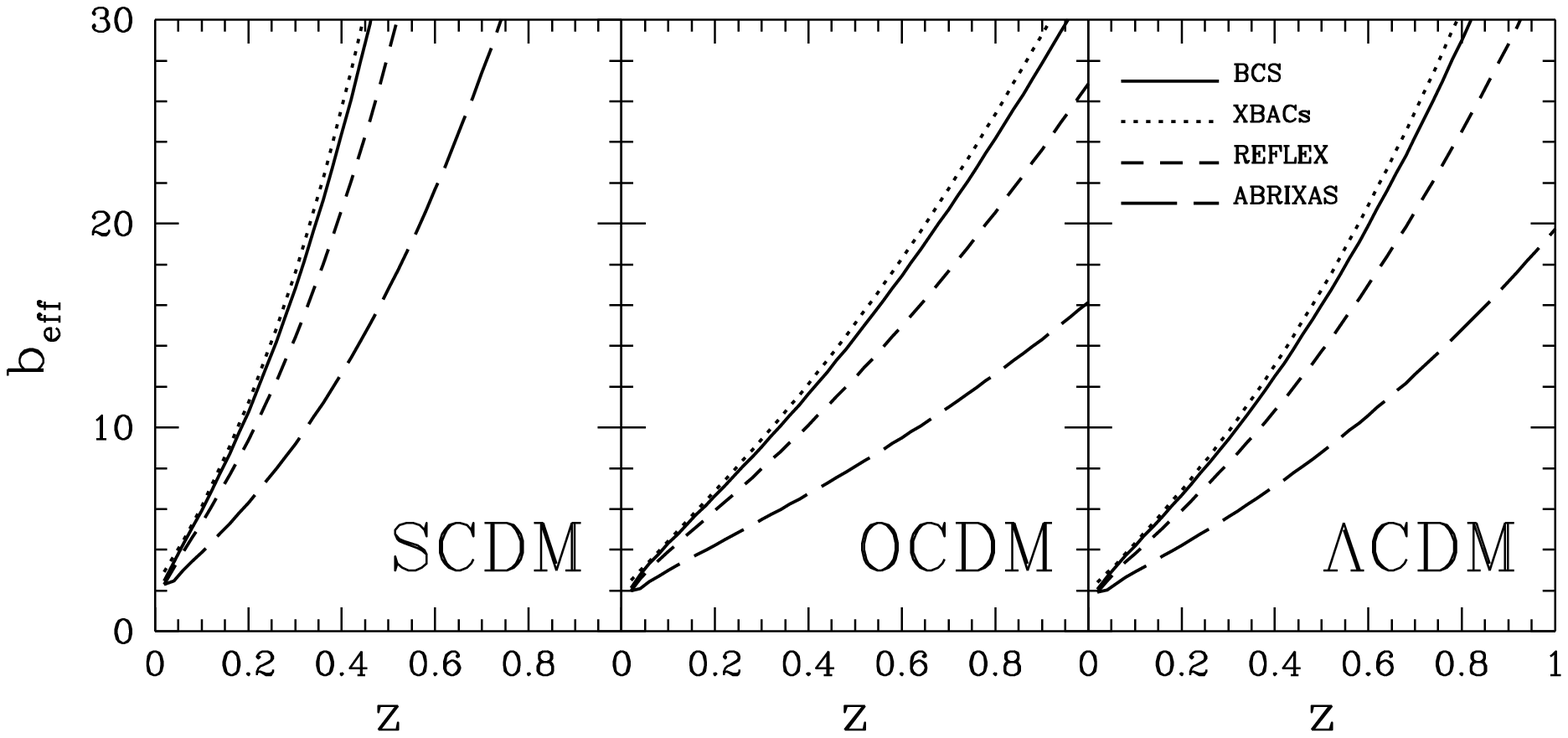,height=8.cm,width=16cm,angle=0}
\caption{The behaviour of the effective bias $b_{\rm eff}$ 
for the clusters included in the flux-limited samples as a function of
the redshift $z$.  Different lines refer to results for different
catalogues: BCS (solid line), XBACs (dotted line), REFLEX (short
dashed) and ABRIXAS (long dashed). Different panels show the
predictions for various cosmological models: SCDM (left), OCDM
(centre) and $\Lambda$CDM (right).}
\label{fi:bias}
\end{figure*}

The last ingredient of our model is the redshift evolution of the
effective bias $b_{\rm eff}(z)$, computed from eq.(\ref{eq:b_eff}). In
Figure~\ref{fi:bias} we show the values of $b_{\rm eff}(z)$ for the
different catalogues. The effective bias is found to be an increasing
function of redshift: high-redshift clusters, if existing, have a very
high bias.

\subsection {Clustering predictions} 

\begin{figure*}
\centering  
\psfig{figure=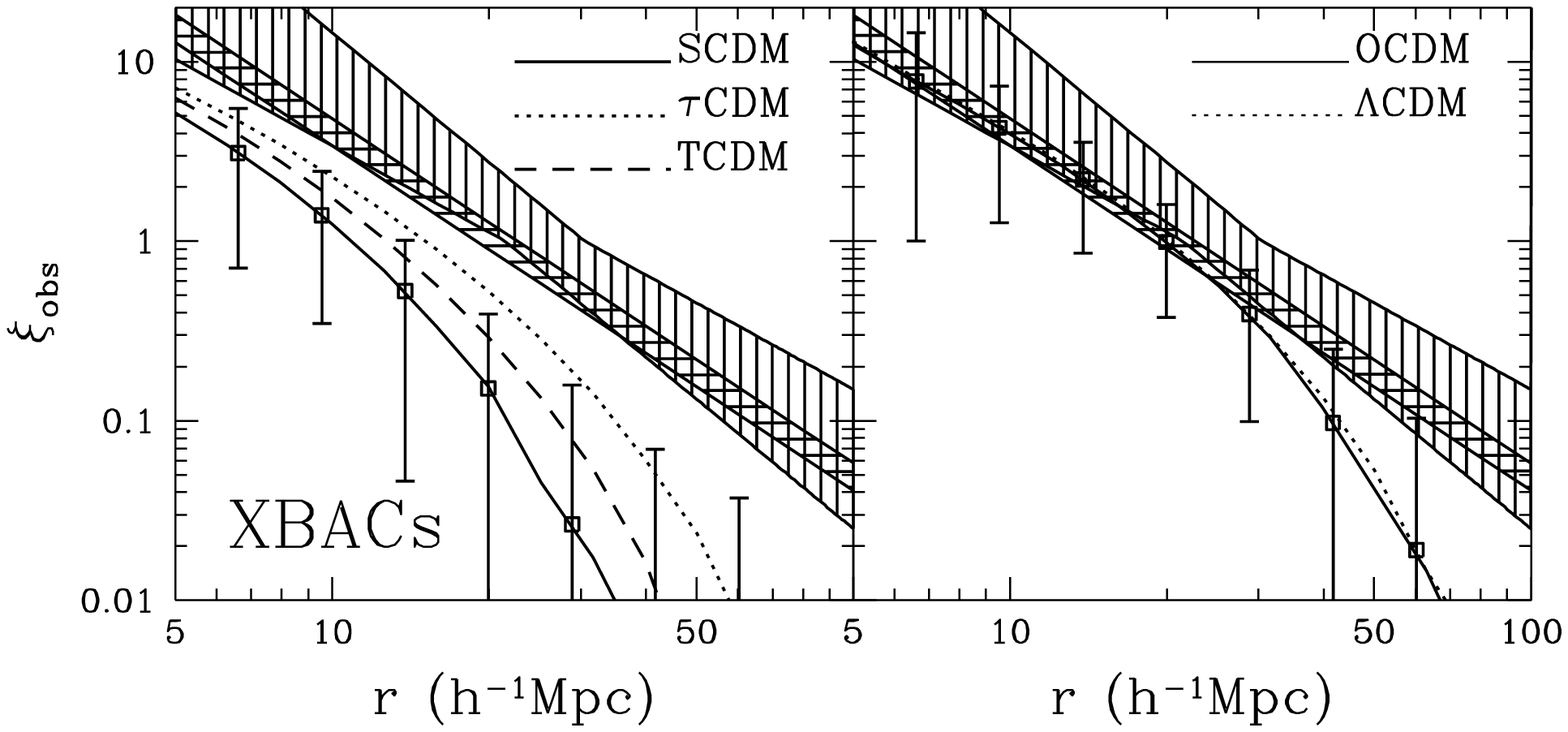,height=8.cm,width=16cm,angle=0}
\caption{Comparison of the observed spatial correlation for clusters in
the XBACs sample with the predictions of the various cosmological
models.  The observational results are shown by the shaded areas: the
horizontal shaded region refers to the (1-$\sigma$) estimates obtained
by Abadi, Lambas \& Muriel (1998), the vertical shaded one shows the
(2-$\sigma$) estimates by Borgani, Plionis \& Kolokotronis (1999).  In
the left panel we present the SCDM model (solid line), the $\tau$CDM
model (dotted line) and the TCDM model (short-dashed line); in the
right panel we show the OCDM model (solid line) and the $\Lambda$CDM
model (dotted line).  Bootstrap estimates of the errorbars
(1-$\sigma$) for the theoretical predictions are shown only for the
SCDM and OCDM models for clarity.  }
\label{fi:xbacs}
\end{figure*}

We start by applying our method to the XBACs catalogue, because in
this case we can compare directly our predictions to the observational
clustering properties obtained by two different groups. Abadi, Lambas
\& Muriel (1998) found that the XBACs spatial correlation function can
be fitted by the usual power-law relation $\xi(r)=(r/r_0)^{-\gamma}$
with $\gamma=1.92$ and $r_0=21.1^{+1.6}_{-2.3}\ h^{-1}$ Mpc (errorbars
correspond to 1 $\sigma$). Borgani, Plionis \& Kolokotronis (1999),
who adopted an analytical approximation to the bootstrap errors for
the variance of $\xi(r)$, found $\gamma=1.98^{+0.35}_{-0.53}$ and a
slightly larger value of $r_0=26.0^{+4.1}_{-4.7}\ h^{-1}$ Mpc
(errorbars in this case are 2-$\sigma$
uncertainties). Figure~\ref{fi:xbacs} compares these observational
estimates (shown by the shaded regions) to the theoretical predictions
of the cosmological models here considered: the three $\Omega_{\rm
0m}=1$ models (SCDM, $\tau$CDM and TCDM) are presented in the left
panel while the two $\Omega_{\rm 0m}=0.3$ models (OCDM and
$\Lambda$CDM) are in the right one.  In the plot `mock-observational'
errorbars are reported only for SCDM and OCDM models, for clarity:
these are obtained by bootstrap resampling the number of expected
pairs in each separation bin (Mo, Jing \& B\"orner 1992).  We find
that all Einstein-de Sitter models display too small correlations.
Their correlation lengths are smaller than the observational results:
we find $r_0\simeq 11, 15, 13\ h^{-1}$ Mpc for SCDM, TCDM and $\tau
CDM$, respectively.  On the contrary, both the OCDM and $\Lambda$CDM
models give very similar results and are in better agreement with the
observational data ($r_0 \simeq 20-22 \ h^{-1}$ Mpc).  Similar
conclusions have been reached by Moscardini et al. (2000) from the
analysis of the RASS1 Bright Sample.

\begin{figure*}
\centering  
\psfig{figure=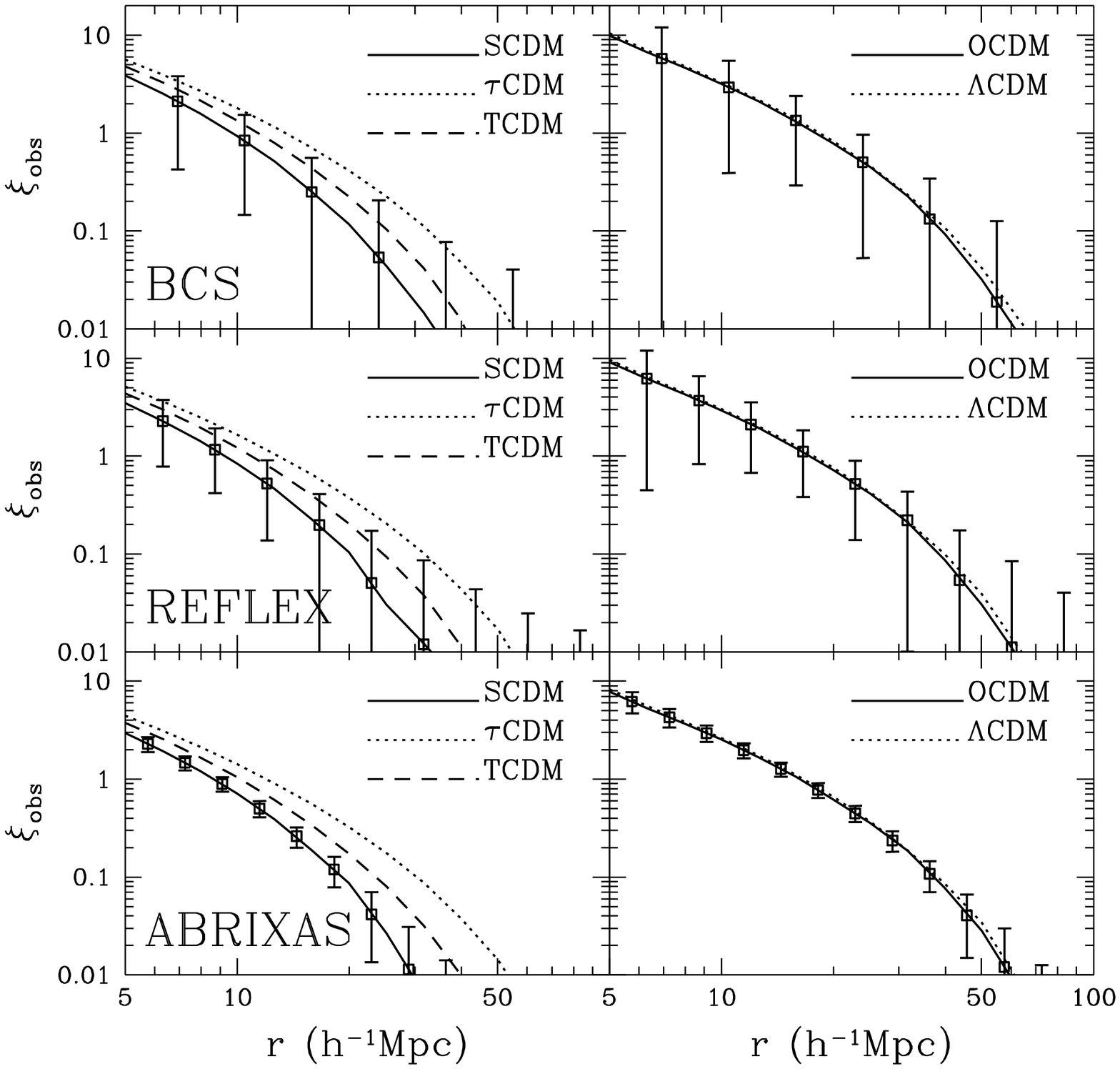,height=16.cm,width=16cm,angle=0}
\caption{Predictions for the spatial correlation of X-ray selected
clusters in the different samples.  Results refer to the BCS sample
(upper row), the REFLEX sample (central row) and the ABRIXAS sample
(lower row).  Different cosmological models are considered.  In the
left column we present the SCDM model (solid line), the $\tau$CDM
model (dotted line) and the TCDM model (short-dashed line); in the
right column we show the OCDM model (solid line) and the $\Lambda$CDM
model (dotted line). Bootstrap estimates of the errorbars (1-$\sigma$)
for the theoretical predictions are shown only for the SCDM and OCDM
models for clarity.}
\label{fi:corre}
\end{figure*}

The theoretical predictions for the other catalogues here considered
(BCS, REFLEX and ABRIXAS) are shown in Figure~\ref{fi:corre}.  Our
results show that the amplitude of the spatial correlation function is
an increasing function of the limiting flux. This is in agreement with
what found by Moscardini et al. (2000), using the RASS1 Bright Sample
data alone, and by Suto et al. (2000) in a more general analysis of
flux-limited surveys made with a technique similar to that applied
here. Moreover, we find that the two non-Einstein-de Sitter models
with $\Omega_{\rm 0m}=0.3$ display similar clustering properties and
tend to have larger amplitudes than the $\Omega_{\rm 0m}=1$
models. Note that preliminary analyses of the two-point spatial
correlation and of the power-spectrum of the REFLEX sample (Collins et
al. 2000; Guzzo et al. 1999; Schuecker et al. 2000), lead to a
correlation length $r_0\simeq 18 \ h^{-1}$ Mpc. If this result will be
confirmed, the comparison with our theoretical predictions will give
further support to cosmological models with a low matter density
parameter.

\begin{figure*}
\centering  
\psfig{figure=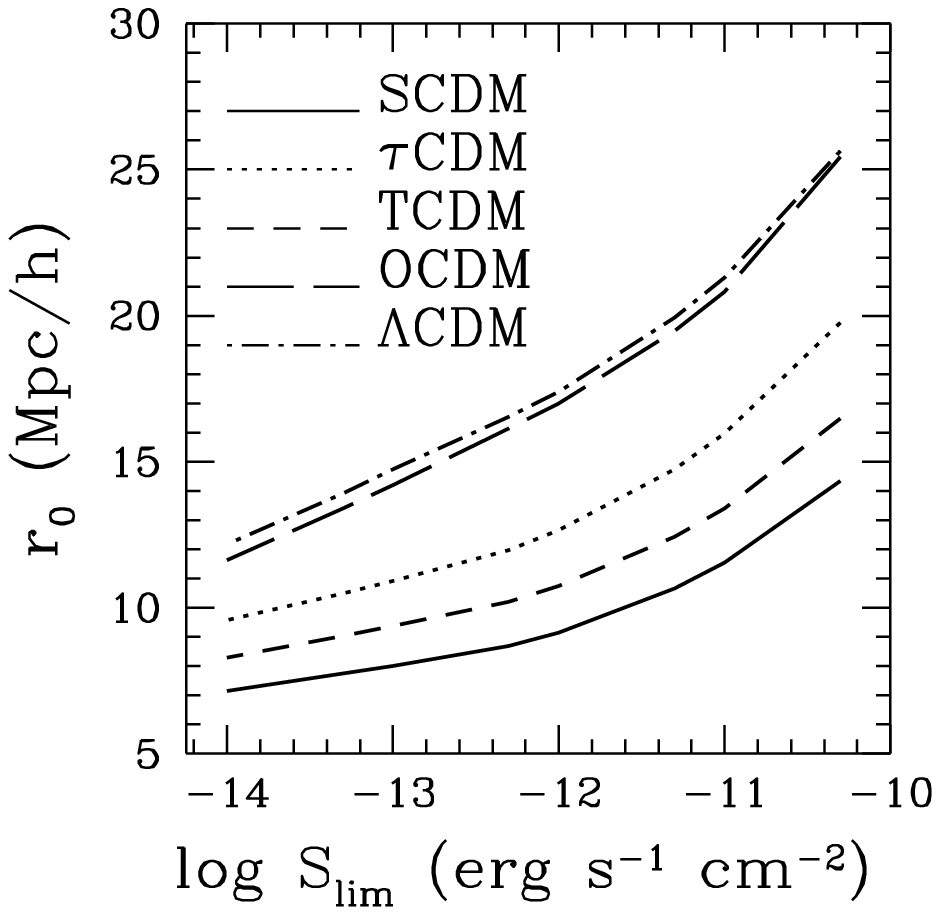,height=8.cm,width=16cm,angle=0}
\caption{The behaviour of the correlation length $r_0$ as a function of 
the limiting X-ray flux $S_{\rm lim}$.  The predictions of the various
theoretical models are shown: SCDM model (solid line), $\tau$CDM model
(dotted line), TCDM model (short-dashed line), OCDM model (long-dashed
line) and $\Lambda$CDM model (dotted-dashed line).}
\label{fi:r0_slim}
\end{figure*}

From the previous analysis we found that the amplitude of $\xi(r)$
decreases by lowering the limiting flux.  This result can be related
to the study of the richness dependence of the cluster correlation
function. In fact, when we change the observational limits, the
resulting sample can have a different mean intercluster separation
$d_c$. A recent analysis of the APM clusters made by Croft et
al. (1997) found a $r_0-d_c$ dependence which is milder than the
linear relation obtained by Bahcall \& West (1992) for the Abell
clusters.  In order to give a more quantitative estimate of this
dependence, we use our model to predict the value of the correlation
length $r_0$ in catalogues where we vary the limiting X-ray flux
$S_{\rm lim}$ (defined in the energy band 0.5--2 keV).  The results,
displayed in Figure~\ref{fi:r0_slim}, confirm that for all the
cosmological models the correlation length $r_0$ grows with $S_{\rm
lim}$.  By changing the limiting flux by four orders of magnitude
(from $S_{\rm lim}= 10^{-14}$ to $S_{\rm lim}=10^{-10}$ erg s$^{-1}$
cm$^{-2}$), the correlation length varies by a factor of $\simeq 2$
(from $r_0 \simeq 7-10$ to $r_0 \simeq 15-20 h^{-1}$ Mpc for the
Einstein-de Sitter models and from $r_0\simeq 12$ to $r_0\simeq 25
h^{-1}$ Mpc for the $\Omega_{\rm 0m}=0.3$ models).

A similar analysis has been made by Suto et al. (2000, see their
Figure 8).  Even if the results cannot be directly compared because
they adopt cosmological models with different values of the parameters
($\Omega_{\rm 0m}$, $\Gamma$ and $\sigma_8$), a different (but almost
equivalent) formalism for the past-light cone effect and different
bias prescriptions, temperature-luminosity and mass-temperature
relations, it is possible to notice that there is general qualitative
agreement. However, our model tends to predict smaller correlation
lengths (by approximately 30 per cent). As we will discuss in Section
5.1, part of this difference comes from the values of ${\cal B}$ used
in the temperature-luminosity relation [see eq.(\ref{eq:t-l})]: ${\cal
B}=1/3.4$ and $1/3$ in Suto et al. (2000) and in this paper,
respectively.  Moreover, in their approach Suto et al. include a
method to account for redshift-space distortion effects (here not
considered) which tends to increase the correlation estimates (see the
discussion at the end of Section 2.1).

The dependence of the correlation length on the survey flux limit can
be used to give a partial explanation of the difference between the
early results derived by Romer et al. (1994) and those obtained in
more recent analyses (Abadi, Lambas \& Muriel 1998; Borgani, Plionis
\& Kolokotronis 1999; Moscardini et al. 2000). In fact the Romer et
al.' catalogue is deeper ($S_{\rm lim}\simeq 10^{-12}$ erg s$^{-1}$
cm$^{-2}$ in the 0.1 -- 2.4 keV band) than both XBACs and the RASS1 Bright
Sample. However, we have also to remind that the results of Romer et
al. (1994) can be affected both by the absence of a study of the
sample sky coverage and by imcompleteness effects. In fact the cluster
catalogue was derived drawing on X-ray information from the ROSAT
standard analysis software (SASS), which was not optimized for the
analysis of extended sources, as shown by De Grandi et al. (1997).

\begin{figure*}
\centering  
\psfig{figure=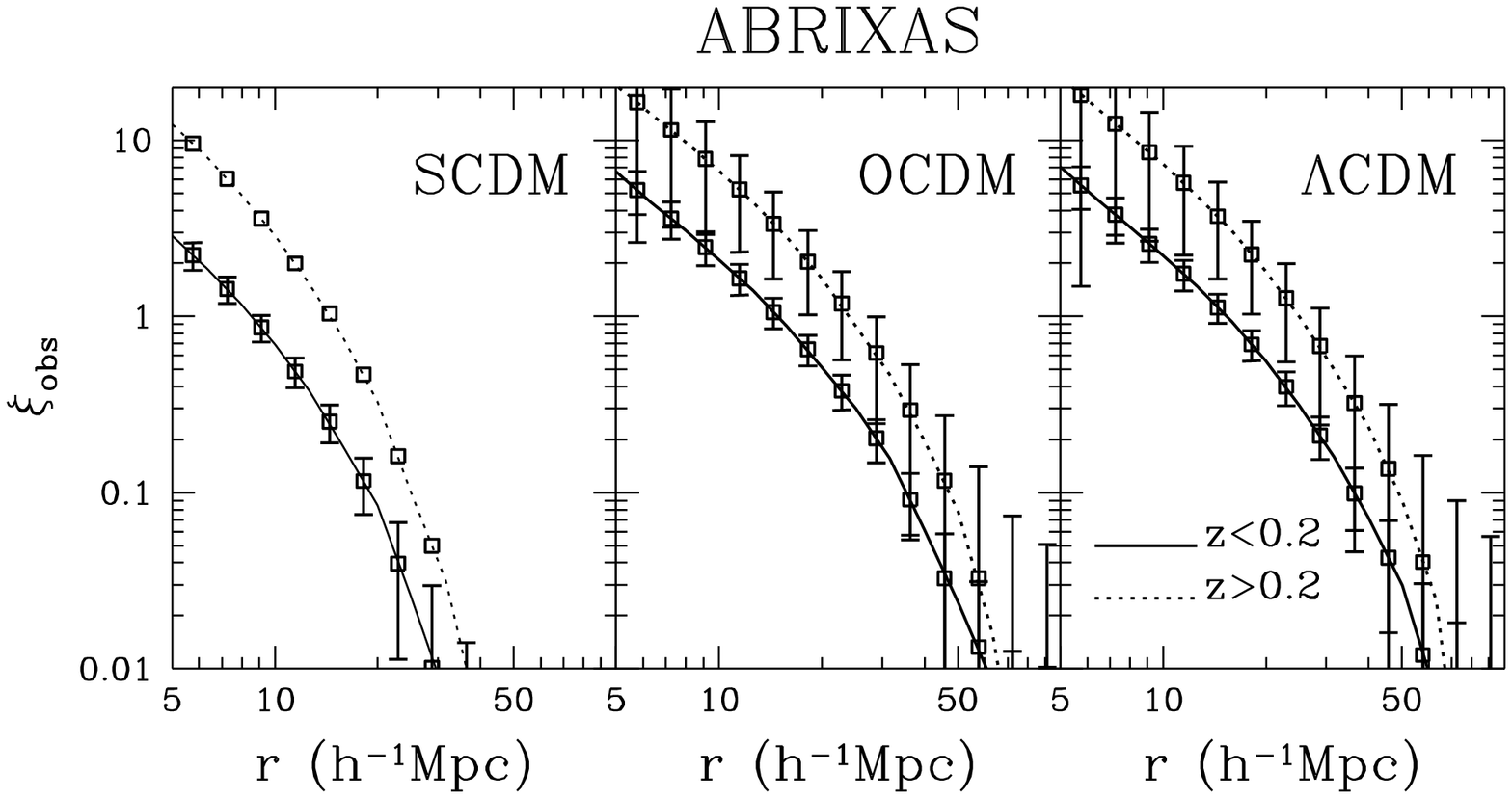,height=8.cm,width=16cm,angle=0}
\caption{Predictions for the redshift evolution of the
spatial correlation $\xi(r)$ of X-ray selected clusters.  Results
refer to two subsamples of the ABRIXAS catalogue: $z<0.2$ (solid line)
and $z>0.2$ (dotted line).  Different cosmological models have been
considered: SCDM (left panel), OCDM (central panel) and $\Lambda$CDM
(right panel).  Errorbars are 1-$\sigma$ bootstrap estimates: they are
not shown for the $z>0.2$ subsample in the SCDM model, because they
are very large due to the small number of predicted objects.}
\label{fi:abr0002}
\end{figure*}

By using a survey as deep and extended as that which was planned with
the ABRIXAS satellite, it would be possible to address the problem of
the redshift evolution of the spatial correlation function
$\xi(r)$. In Figure~\ref{fi:abr0002} we show the theoretical
predictions for $\xi(r)$ obtained if the ABRIXAS catalogue is divided
in two subsamples, $z<0.2$ and $z>0.2$.  All cosmological models
display a larger amplitude at higher redshifts.  The difference
between the $z<0.2$ and $z>0.2$ subsamples is significant: the
correlation length moves from $r_0 \simeq 8$ to $r_0\simeq 15 h^{-1}$
Mpc for the SCDM model and $r_0 \simeq 15$ to $r_0\simeq 25 h^{-1}$
Mpc for the OCDM and $\Lambda$CDM models.  Notice that the errorbars
for the high-redshift subsample in the SCDM model are not shown
because they are very large due to the small number of expected
objects.

\begin{figure*}
\centering  
\psfig{figure=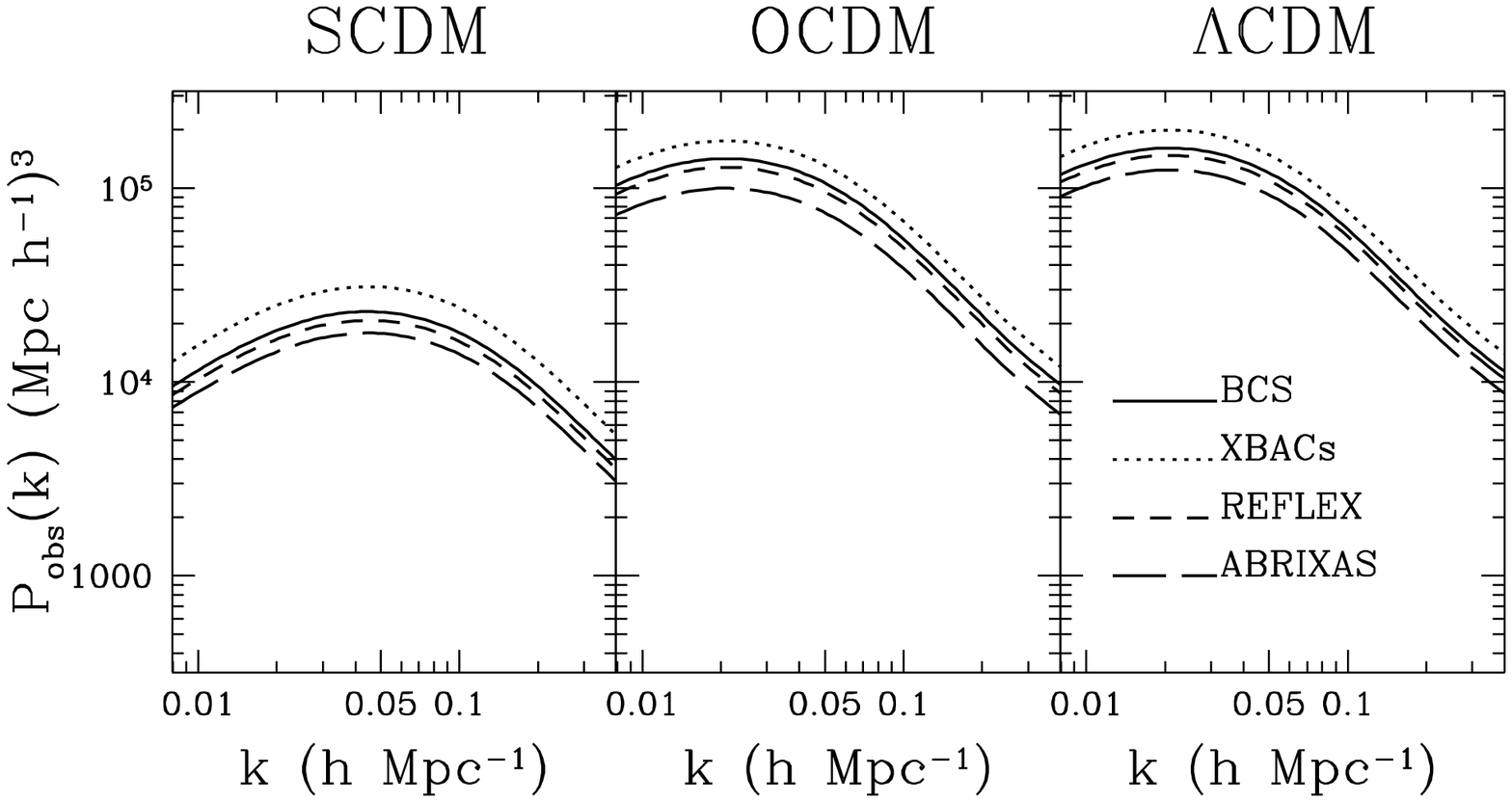,height=8.cm,width=16cm,angle=0}
\caption{The theoretical predictions for the power-spectrum as
measured in various catalogues: BCS (solid line), XBACs (dotted line),
REFLEX (short dashed) and ABRIXAS (long dashed). Different panels show
the predictions for various cosmological models: SCDM model (left),
OCDM model (centre) and $\Lambda$CDM model (right).}
\label{fi:power}
\end{figure*}

The power-spectrum analysis is an alternative way to study the
clustering properties of a given catalogue.  However, up to now this
technique has been very seldom applied to X-ray selected clusters
because of the difficulties to correctly take into account the actual
sky coverage (see however Schuecker et al. 2000). By using our
formalism [see eq.(\ref{eq:pow2})] we can obtain the expected
power-spectrum $P_{\rm obs}(k)$ measured by the different surveys.
The results for SCDM, OCDM and $\Lambda$CDM are shown in
Figure~\ref{fi:power}.  As it is possible to notice, given a
cosmological model, the shape of $P_{\rm obs}(k)$ does not change by
varying the limiting flux: the unique effect is the change of the
amplitude which decreases when $S_{\rm lim}$ decreases.

To conclude this subsection we would like to stress the relevance of
light-cone effects in the present problem. Several authors have
estimated the spatial clustering of X-ray clusters in flux-limited
surveys using either analytical treatments or N-body simulations at a
fixed redshift (typically chosen to coincide with the median redshift
of the catalogue).  The level of inaccuracy inherent in such an
approach, even for shallow surveys, has been already discussed in
(Moscardini et al. 2000), who found that for the RASS1 Bright sample
the traditional method would lead to a 20 per cent overestimate of the
correlation length. For deeper surveys, such as ABRIXAS, we find that
the error would be even larger, reaching approximately 25 per cent in
the case of the cosmological models with $\Omega_{\rm 0m}=0.3$.  A
somewhat better approximation would be obtained by considering
constant-time simulations (or analytical expressions for the two-point
function) at an ``effective redshift'' $z_{\rm eff}$, defined as the
peak of ${\cal N}^2(z)/g(z)$ of the sample [see eq.(\ref{eq:xiobs2})].

\subsection {Strategy for future surveys}

Another possible application of our technique is in the preparation of
the strategy for future surveys.  A typical problem to be addressed is
the choice of the limiting flux to be reached (related to a minimum
number of counts) and of the size of the sky area $\cal A$ to be
covered.  Of course, the best solution would be to have small $S_{\rm
lim}$ and large $\cal A$, but, given a finite life-time for an X-ray
satellite, the two choices are in competition.  To discuss this
problem, we will consider two different surveys with characteristics
already possible with the presently proposed space missions (see
e.g. Chincarini 1999).  The first one is meant to represent an example
of a survey covering a wide area but with a fainter limiting flux
(hereafter called `WIDE'); the second one (hereafter `DEEP') simulates
a very deep survey but with a more limited sky coverage. We choose as
characteristic parameters $S_{\rm lim}=3 \times 10^{-14}$ erg
cm$^{-2}$ s$^{-1}$ and ${\cal A}=1000$ deg$^2$ to define the WIDE
catalogue, and $S_{\rm lim}= 7 \times 10^{-15}$ erg cm$^{-2}$ s$^{-1}$
and ${\cal A}= 100$ deg$^2$ for the DEEP one (the fluxes are in the
0.5--2 keV band).  The predictions for both the spatial and angular
correlation functions [computed by using eq.({\ref{eq:ang2})] are
shown in Figure~\ref{fi:future}. They refer only to the OCDM model,
used as a working example.  The important feature of this plot is the
size of the errorbars which are 1-$\sigma$ bootstrap estimates. For
the WIDE survey the clustering signal is larger than the errorbars and
can be detected also for scales larger than 2000 arcsec and $30\
h^{-1}$ Mpc, for $\omega_{\rm obs}$ and $\xi_{\rm obs}$
respectively. On the contrary, for the DEEP survey the angular
correlation has a significant signal only up to $\simeq 500$ arcsec
while the spatial correlation is completely dominated by the noise on
all scales. From these results we can conclude that the best strategy
to study the clustering properties should be based on wide angle,
relatively shallow, surveys.

\begin{figure*}
\centering  
\psfig{figure=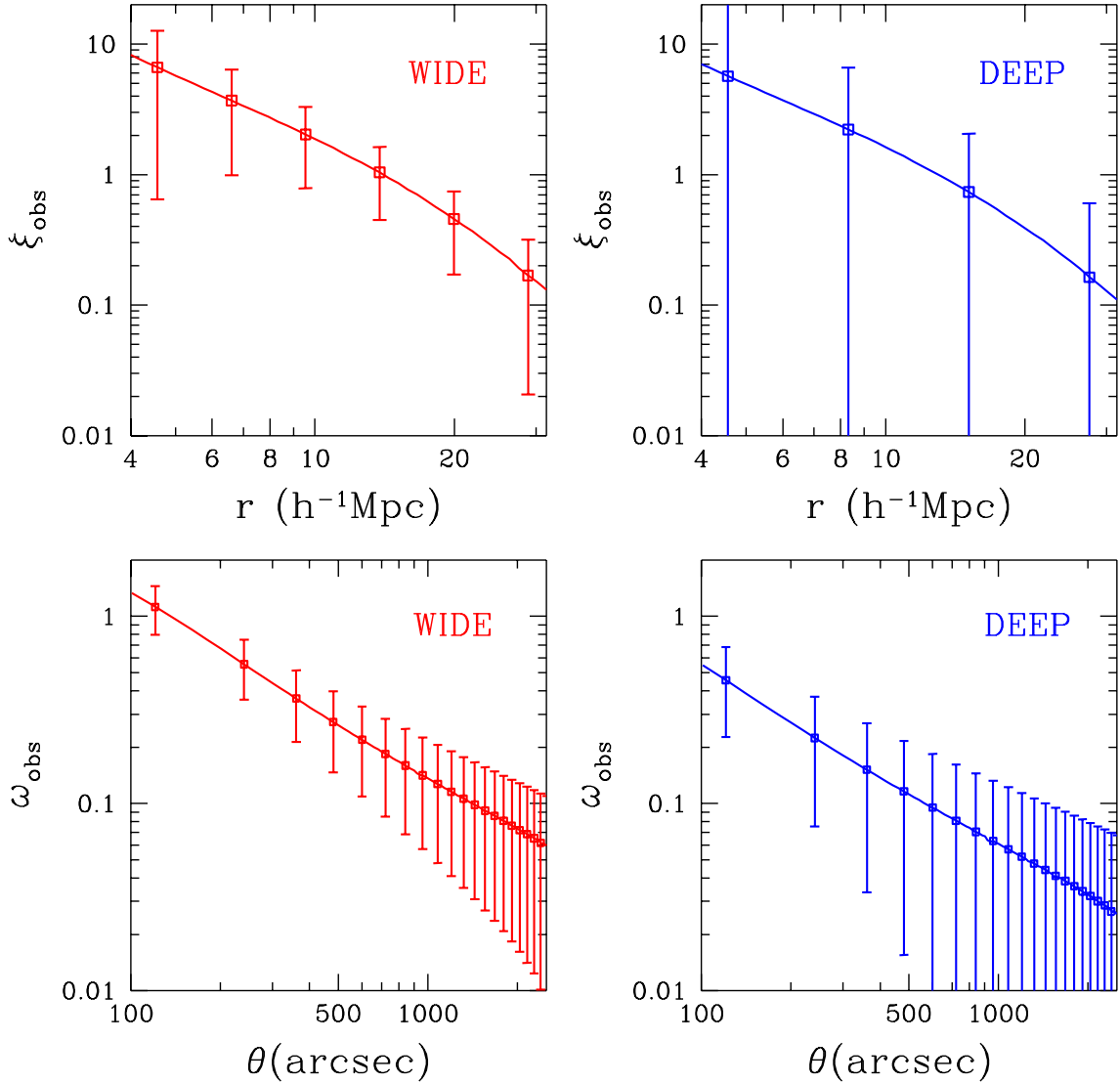,height=16.cm,width=16cm,angle=0}
\caption{Theoretical predictions for the spatial correlation function
$\xi_{\rm obs}$ (upper panels) and for the angular correlation
function $\omega_{\rm obs}$ (lower panels) for the WIDE ($S_{\rm
lim}=3 \times 10^{-14}$ erg cm$^{-2}$ s$^{-1}$ and ${\cal A}=1000$
deg$^2$) and DEEP ($S_{\rm lim}= 7 \times 10^{-15}$ erg cm$^{-2}$
s$^{-1}$ and ${\cal A}= 100$ deg$^2$) surveys, shown in the left and
right columns, respectively.  Results are obtained using the OCDM
model.  Errorbars are 1-$\sigma$ bootstrap estimates.}
\label{fi:future}
\end{figure*}
 
\section{Robustness of the results}
\subsection{Dependence on the temperature-luminosity relation}

Our model to predict the clustering properties of X-ray selected
galaxy clusters makes use of various relations to translate the
limiting flux (which is the observational quantity) into the minimum
mass of the hosting dark matter halo (which is the variable preferred
in the theory). The largest uncertainties inherent in this approach
are in the temperature-luminosity relation. As mentioned in Section
3.4, to this aim we assumed the semi-empirical power-law relation
$T\propto L_{\rm bol}^{\cal B}$, with ${\cal B}=1/3$.  Moreover, we
fixed a minimum temperature of 1 keV and we allowed a mild redshift
evolution of the $T-L_{\rm bol}$ relation to reproduce the cluster
number counts.  In this subsection we will briefly discuss the
stability of the previous results with respect to changes in these
choices. For that we will consider only results for the ABRIXAS survey
since it is the deepest one and consequently the possible effects will
be enhanced.

As already said, the value of $\cal B$ we used is in agreement with
various local (i.e. at $z\simeq 0 $) estimates (e.g. David et
al. 1993; White, Jones \& Forman 1997; Markevitch 1998).  However,
different values of $\cal B$ could be considered. The scaling
relations, for example, suggest ${\cal B}=1/2$; hydrodynamical
simulations without inclusion of heating and cooling effects give
support to this value, while smaller values of $\cal B$, more similar
to the observational estimates, are obtained when the supernova
feedback is included (e.g. Navarro, Frenk \& White 1995; Cavaliere,
Menci \& Tozzi 1999). One more systematic effect can be due to the
presence of cooling flows in the central part of the
clusters. Corrections for this effect have been tried resulting in an
increase of the observed $\cal B$ (e.g. White, Jones \& Forman 1997;
Allen \& Fabian 1998; Arnaud \& Evrard 1999), in better agreement with
the scaling relation.
  
To study the effect of this uncertainty we allow the parameter $\cal
B$ to change in the range $1/3.5 \le {\cal B} \le 1/2.5$. The results
are shown in Figure~\ref{fi:test} for SCDM, OCDM and
$\Lambda$CDM. Notice that we are still imposing a minimum temperature
of 1 keV and requiring a redshift evolution of the $T-L_{\rm bol}$
relation to reproduce the counts: the resulting values of the $\eta$
parameter [see eq.(\ref{eq:t-l})] become larger (smaller) when the
value of $\cal B$ is decreased (increased). From the plot it is
possible to see that by varying the $\cal B$ parameter the resulting
spatial correlation function does not change its shape, but only the
amplitude: the larger is $\cal B$, the smaller is $\xi(r)$. However,
the changes are quite small and with a size similar to the expected
bootstrap errorbars (see e.g. Figure~\ref{fi:corre}).

\begin{figure*}
\centering  
\psfig{figure=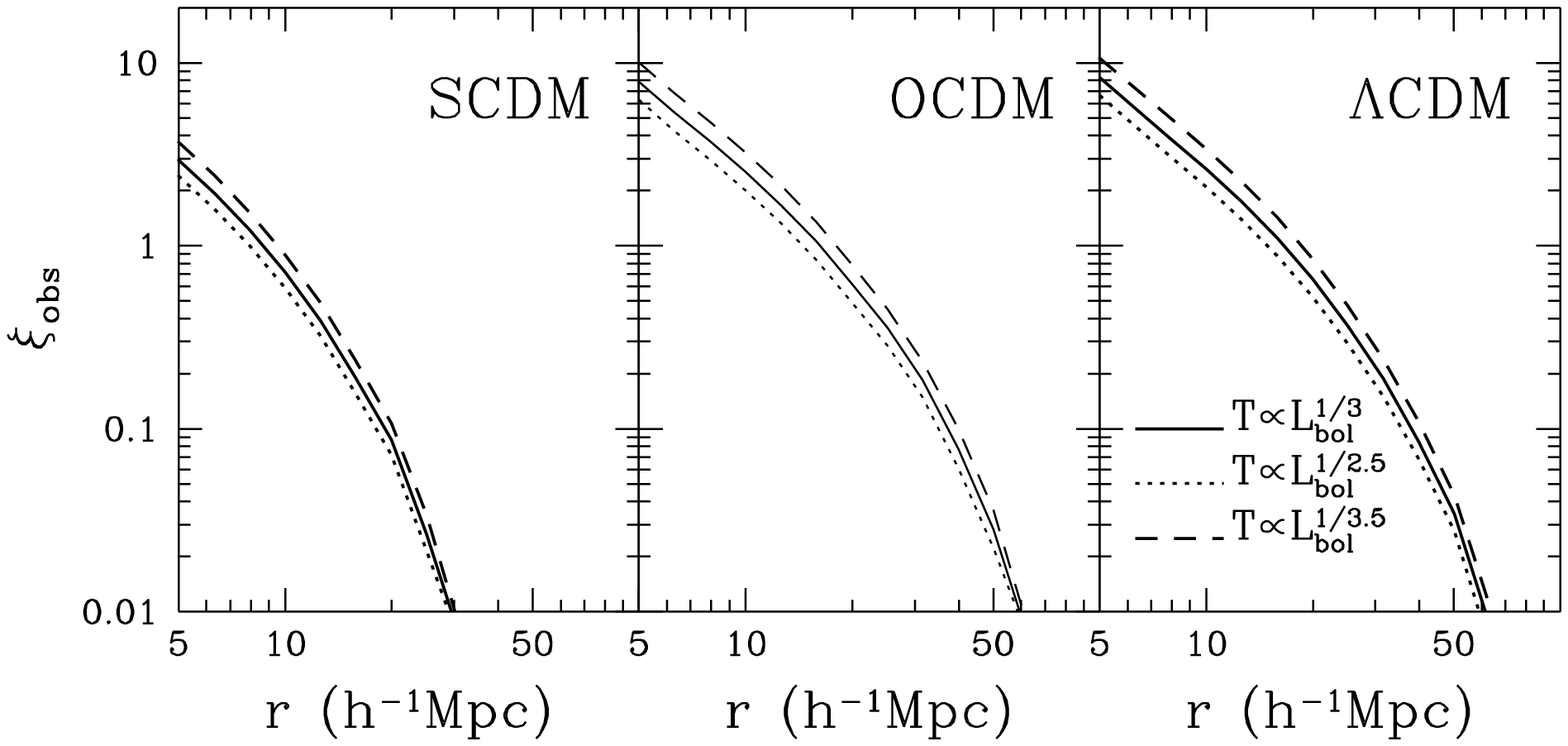,height=8.cm,width=16cm,angle=0}
\caption{Theoretical predictions for the spatial correlation function
$\xi_{\rm obs}$ estimated assuming different temperature-luminosity
relations $T\propto L_{\rm bol}^{\cal B}$: solid, dotted and dashed
lines refer to ${\cal B}= 1/3, 1/2.5$ and 1/3.5, respectively.
Results refer to the ABRIXAS catalogue and are obtained for the SCDM
model (left panel), the OCDM model (central panel) and the
$\Lambda$CDM model (right panel).}
\label{fi:test}
\end{figure*}

We also considered the effect of including clusters with temperature
$T<1$ keV (excluded in the previous analyses) adopting, only at small
$T$, the $T-L_{\rm bol}$ relation obtained by Ponman et al. (1996)
from an analysis of groups of galaxies.  This corresponds to a small
value of the slope (${\cal B}=1/8.2$). The resulting change in the
estimates of the spatial correlation function (not shown in the plot)
are negligible and always smaller than those previously obtained. A
similar result is also obtained in two other cases: when we remove the
constraint on the minimum temperature and when we avoid the redshift
evolution of the $T-L_{\rm bol}$ relation (i.e. when $\eta=0$).

\subsection{Dependence on the normalization of the cosmological models}

In this paper we normalized the cosmological models here considered by
using the local cluster abundance. In particular we adopted the values
of $\sigma_8$ coming from the Eke, Cole \& Frenk (1996) analysis of
the temperature distribution of X-ray clusters. Strictly speaking,
this approach would be appropriate only for clusters having a typical
temperature of $T\sim 5-6$ keV, while in our analysis we consider
clusters down to the imposed cutoff of $T=1$ keV.  One of the
consequences of our choice is that the number of clusters predicted
for SCDM is larger than that expected for the low-density models (see
our Figure~\ref{fi:nz} and Figure 5 of Suto et al. 2000).  Since the
cosmological models are all normalized to the local cluster abundance
and this abundance declines more rapidly with increasing redshift when
$\Omega_0$ is high, one would expect the opposite trend.  The
explanation is that SCDM has a larger number of relatively low
temperature ($T\sim 1$ keV) clusters with respect to the other models
because of its steeper mass (and temperature) function.  This feature
could introduce a dependence of some results on the imposed lower
temperature.

A way to overcome the problem could be to normalize the cosmological
models by using the local X-ray luminosity function. In fact, unlike
the temperature function, it extends to clusters with $T\sim 1$ keV.
Borgani et al. (1999) followed this approach using the luminosity
functions coming from the RDCS (Rosati et al. 1998) and BCS (Ebeling
et al. 1997) samples.  The result is a relation between $\sigma_8$ and
the density parameter $\Omega_0$ corresponding to the best-fit of the
data to the model predictions.  In order to study the dependence of
our results on different choices for the normalization, we computed
the clustering properties for the ABRIXAS case adopting the values of
$\sigma_8$ resulting from the Borgani et al. (1999) relation, namely
$\sigma_8=0.58$, 1.00 and 0.96 for SCDM, OCDM and $\Lambda$CDM models,
respectively. The spatial correlation functions (not reported here)
show differences smaller than 10 per cent with respect to those
obtained with the previous normalizations.  Similar differences are
obtained for the other catalogues here considered.

\section{Conclusions} 

The purpose of this paper was to present predictions for the
clustering properties of X-ray selected clusters as measured in
flux-limited surveys. To this aim we introduced a model which accounts
for the clustering of observable objects in our past light-cone and
for the redshift evolution of both the underlying dark matter
covariance function and the cluster bias factor. Our approach makes
use of theoretical and empirical relations between mass, temperature
and X-ray luminosity of galaxy clusters which allow to translate the
limiting flux of a survey (which is the observational quantity) into a
corresponding minimum mass for the dark matter haloes hosting the
clusters. The results of the application of this method have been
found to be only slightly sensitive to the power-spectrum
normalization of the cosmological models and to the parameters
entering in the adopted semi-empirical relations (within a sensible
range).

The model, which is able to reproduce the observed cluster counts
($\log N-\log S$) by allowing a mild redshift evolution of the
temperature-luminosity relation, has been applied to obtain
predictions for the two-point (spatial and angular) correlation
functions and power-spectrum for present and future surveys in the
framework of different cosmological scenarios. Our main results are as
follows:
\begin{itemize}
\item 
In surveys with different limiting flux, the observed clusters have
different properties in terms of luminosity, temperature and mass. In
particular, reducing the limiting fluxes corresponds to adding smaller
mass haloes.  Since the bias factor is an increasing function of the
mass, we find that the amplitude of the spatial correlation function
of X-ray selected clusters decreases when the limiting flux is
lowered. A similar conclusion has been very recently reached by Suto
et al. (2000).
\item
We made predictions concerning different catalogues (BCS, XBACs and
REFLEX) by using the actual sky coverage where available (REFLEX) and
by correcting for incompleteness where necessary (XBACs). The results
show that the Einstein-de Sitter models here considered always give
smaller correlation amplitude compared to models with low matter
density parameter, $\Omega_{\rm 0m}=0.3$.
\item
In the case of the XBACs catalogue, it is possible to compare our
predictions with two different observational estimates (Abadi, Lambas
\& Muriel 1998; Borgani, Plionis \& Kolokotronis 1999). As already
found in a similar analysis of the RASS1 Bright Sample by Moscardini
et al. (2000), only the models with low $\Omega_{\rm 0m}$ can
reproduce the observed clustering, while the correlation strength for
the Einstein-de Sitter models is too low.
\item
We applied our method also to make predictions for possible future
surveys of X-ray clusters. Our results show that in subsamples of
high-redshift objects the amplitude of the correlation function would
be higher than the local one.
\item
Finally, we used our approach to discuss what would be the best
strategy for future surveys, by comparing the clustering signal
obtained with two different choices: a deep survey on a small area of
the sky vs. a brighter survey covering a wider area.  We found that
the second configuration is preferable because of the size of
the errorbars which would allow the detection of the clustering (both
spatial and angular) on relevant scales.
\end{itemize}

In conclusion, we think that the clustering properties of X-ray
selected clusters are a very powerful tool to study the large-scale
structure of the universe. In fact, our results show that this
approach, similarly to the study of the cluster abundances, can be
successfully used to put constraints on the cosmological parameters.
Such a method will become more powerful when the data for deeper
surveys will be available in the next future.

\section*{Acknowledgments.} 

This work has been partially supported by Italian MURST, CNR and
ASI. We want to warmly thank Sabrina De Grandi for stimulating
comments and helpful suggestions and for giving us her results about
number counts at high limiting flux. We are grateful to Hans
B\"ohringer and C. Collins for having provided the sky coverage of the
REFLEX catalogue and to Stefano Borgani, Ornella Pantano, Yasushi Suto
and Bepi Tormen for useful discussions.  We also thank the referee,
Vincent Eke, for comments which allowed us to improve the presentation
of this paper.

\end{document}